\documentclass[journal,twocolumn]{IEEEtran}
\usepackage{epsfig,makeidx,color}
\usepackage{amsmath,amssymb,amsthm,bbm}
\usepackage{cite,graphicx}
\usepackage{enumerate}
\usepackage{hyperref}
\hypersetup{
        colorlinks = true,
        citecolor=blue,
}

\def\cL{{\cal L}}

\def\cH{{\cal H}}

\def\rH{{\rm H}}

\def\rT{{\rm T}}

\def\uC{{\mathbb C}}
\def\uE{{\mathbb E}}

\DeclareMathOperator*{\argmin}{\arg\!\min}
\DeclareMathOperator*{\argmax}{\arg\!\max}

\newtheorem{mylemma}{\bf Lemma} 

\def\be{ \begin{equation} }
\def\ee{ \end{equation} }
\def\bea{ \begin{eqnarray} }
\def\eea{ \end{eqnarray} }
\def\bx{{\bf x}}
\def\by{{\bf y}}

\def\bd{{\bf d}}

\def\bs{{\bf s}}
\def\ba{{\bf a}}
\def\br{{\bf r}}
\def\bu{{\bf u}}

\def\bn{{\bf n}}

\def\bz{{\bf z}}

\def\bv{{\bf v}}

\def\bA{{\bf A}}

\def\bF{{\bf F}}

\def\bH{{\bf H}}
\def\bI{{\bf I}}

\def\bR{{\bf R}}
\def\bW{{\bf W}}

\def\bX{{\bf X}}

\def\b0{{\bf 0}}

\def\bPsi{{\bf \Psi}}

\def\bpsi{{\boldsymbol \psi}}

\def\cC{{\cal C}}

\def\cR{{\cal R}}

\def\cN{{\cal N}}
\def\cS{{\cal S}}

\ifCLASSOPTIONonecolumn
  \newcommand{\figwidth}{0.60\columnwidth}
\else
  \newcommand{\figwidth}{0.90\columnwidth}
\fi

\begin{document}

\title{Single-Carrier Index Modulation for IoT Uplink}

\author{Jinho Choi
\thanks{The author is with
the School of Information Technology,
Deakin University, Geelong, VIC 3220, Australia
(e-mail: jinho.choi@deakin.edu.au).}}

\IEEEspecialpapernotice{(Invited Paper)} 

\maketitle

\begin{abstract}
For the Internet of Things (IoT), 
there might be a large number of devices to be connected to 
the Internet through wireless technologies.
In general, IoT devices would have various constraints
due to limited processing capability, memory, energy source, and so on,
and it is desirable to employ efficient 
wireless transmission schemes, especially for uplink transmissions.
For example,
orthogonal frequency division multiplexing (OFDM) with 
index modulation (IM) or OFDM-IM 
can be considered for IoT devices due to its energy efficiency.
In this paper, we study a different IM scheme
for a single-carrier (SC) system, which is referred to as SCIM.
While SCIM is similar to OFDM-IM in terms of energy efficiency,
SCIM may be better suited for IoT uplink because 
it has a low peak-to-average power ratio (PAPR)
and does not require inverse fast Fourier transform (FFT) at devices 
compared to OFDM-IM.
We also consider precoding 
for SCIM and generalize it to multiple access channel
so that multiple IoT devices can share the same radio resource block.
To detect precoded SCIM signals, low-complexity detectors are derived.
For a better performance, based on variational inference
that is widely used in machine learning,
we consider a detector that provides an approximate
solution to an optimal detection.
\end{abstract}

{\IEEEkeywords
sparsity; index modulation; compressive sensing; diversity}

\ifCLASSOPTIONonecolumn
\baselineskip 26pt
\fi

\section{Introduction} \label{S:Intro}

The Internet of Things (IoT) has attracted attention 
from both academia and practitioners as it 
can support a number of new services and applications through the network
of various (electronic) devices, sensors, and 
actuators \cite{Atzori10} \cite{ITU_IoT} \cite{Fuqaha15}.
While some IoT devices are connected through wired networks,
the connectivity of most IoT devices (such as
sensors) would rely on wireless technologies \cite{Sharma16}.
For example, ZigBee \cite{Gutierrez03}, 
which has been used for wireless sensor networks (WSNs), can be
employed to support the connectivity of IoT devices.

In cellular systems, machine-type communication (MTC) has been
considered for the connectivity of machines including IoT devices
\cite{3GPP_MTC} \cite{3GPP_NBIoT}.
In particular, narrowband-IoT (NB-IoT)  \cite{3GPP_NBIoT} \cite{Wang17}
is to support a large number of IoT devices in a cell.
NB-IoT is based on Long-Term Evolution (LTE) standards
\cite{Dahlman13}
with a system bandwidth of 180 KHz for each of uplink and downlink.
In particular, orthogonal frequency
division multiple access (OFDMA) is adopted 
for downlink, while single-carrier 
frequency division multiple-access (SC-FDMA) is used for uplink
as in LTE, which allows to reduce 
development time for NB-IoT equipments and products.
In addition,
since most IoT devices have various constraints and limitations,
NB-IoT focuses on low-cost design and high energy efficiency
for IoT devices.

Orthogonal 
frequency division multiplexing
(OFDM) with index modulation (IM) (OFDM-IM)
has been proposed in \cite{Basar13} (see also \cite{Abu09}),
where a subset of subcarriers are active and the indices
of them are also used to convey information bits.
The main advantage of OFDM-IM over conventional OFDM
is the energy efficiency and
robustness against inter-carrier interference (ICI)
because only a fraction of subcarriers are
active \cite{Xiao14b}.
A generalization of OFDM-IM is discussed in \cite{Fan15}
where the number of active subcarriers for each sub-block
is not fixed, but variable to increase the number of bits
to be transmitted.
Furthermore, another scheme that independently applies
IM to in-phase and quadrature phase components of complex
symbols is considered to increase the number of bits per sub-block
in \cite{Fan15}.
In \cite{Ko14}, a performance analysis is carried out
when the maximum likelihood (ML) detector is employed.
An excellent overview of IM techniques 
including spatial modulation \cite{Mesleh08} \cite{Jeganathan08}
\cite{Renzo11}
can be found in \cite{Basar16}.

In OFDM-IM, the set of subcarriers
is divided into multiple subsets or clusters and in each subset
both IM and conventional modulation such as quadrature
amplitude modulation (QAM) are employed 
to transmit information bits \cite{Basar13}. 
In general, the number of subcarriers per cluster
is not large in order to avoid a high computational complexity
for the ML detection, while the number of information bits 
can increase with the size of cluster (for a fixed number of subcarriers).
Thus, there is a trade-off between
the size of cluster (and the number of information bits)
and the receiver complexity in OFDM-IM.
In order to allow a low-complexity detection
without dividing the set of subcarriers into clusters,
sparse IM is considered in \cite{Choi_Ko15}, which is also 
used for multiple access \cite{JChoi15} \cite{Choi16}.
Due to the sparsity of active subcarriers in sparse IM, the notion
of compressive sensing (CS) \cite{Donoho06} \cite{Candes06}
can be exploited to derive low-complexity detection methods.

OFDM-IM can also be applied to multiple input multiple output (MIMO)
systems \cite{Basar15}. In \cite{Zheng17},
low-complexity detection approaches 
are considered for MIMO OFDM-IM. Provided that a transmitter
knows the channel state information (CSI), precoding can be employed for
MIMO OFDM-IM as in \cite{Gao18}, which results in a performance
improvement. Since OFDM-IM has a limited diversity gain
\cite{Basar13}, transmit diversity techniques can be considered
to increase the diversity gain. In \cite{Basar15_CI}, 
a transmit diversity technique based on space-time coding,
is applied to OFDM-IM,
which is called coordinated interleaved OFDM-IM (CI-OFDM-IM),
in order to improve
the diversity gain. Variations of CI-OFDM-IM
to further improve the performance are also studied in \cite{Liu17}
\cite{Li18}. 
Channel coding with repetition diversity 
and trellis coded modulation (TCM) 
are applied to OFDM-IM 
in \cite{Choi17_CIM} 
and \cite{Choi18_TCM}, respectively.

Although OFDM-IM has a high energy efficiency and would be suitable 
for energy limited IoT devices,
it has drawbacks that are inherited from OFDM (e.g., a high
peak-to-average power ratio (PAPR)\footnote{Comparisons 
between SCIM and OFDM-IM in terms of PAPR can be found in \cite{Sugiura17}.}, 
and no path diversity gain for uncoded signals).
In addition, in NB-IoT, since SC-FDMA (not OFDM) is employed 
for IoT uplink as mentioned earlier,
IM schemes for SC-FDMA or single-carrier (SC) systems are desirable.
To this end, single-carrier index modulation or SCIM
studied in \cite{JC_ICC17} \cite{Nakao17}
can be considered for IoT uplink 
with a few advantages of SC system 
over multicarrier (MC) system including
the path diversity gain for uncoded signals \cite{Falconer02}
(as SCIM inherits the advantages of SC (over MC)). 
In particular, as shown in \cite{Sugiura17}, 
the PAPR of SCIM is lower than that of OFDM-IM.
In comparison to OFDM-IM, another prominent feature 
of SCIM in terms of IoT uplink is that 
the complexity of the transmitter in IoT devices 
may be low, since no inverse fast Fourier transform (FFT) is required.
In addition, SCIM can exploit the path diversity gain
and performs better than OFDM-IM,
while the notion of sparse IM can be applied to SCIM
so that low-complexity CS algorithms can be used
for the signal detection \cite{JC_ICC17}.

In this paper, we mainly consider SCIM with precoding
and study different approaches to the signal detection.
In particular, we consider detection approaches
that can exploit the sparsity of signals in SCIM.
For precoding, faster-than-Nyquist (FTN) signaling \cite{Mazo75}
is mainly considered. We also generalize SCIM with precoding to
multiple access channel
so that multiple IoT devices can share the same radio
resource block for uplink transmissions.
Note that FTN signaling has been applied to SCIM in \cite{Ishihara17}. 
Thus, the approach in this paper can be
seen as a generalization of the approach in \cite{Ishihara17} 
in terms of precoding (i.e., FTN is seen as a special case of precoding)
and with multiple access to support
multiple users in the same resource block.

The rest of the paper is organized as follows.
In Section~\ref{S:SM},
we present the system model for SCIM
over an intersymbol interference (ISI) channel.
Precoding is applied to SCIM, which can increase 
the number of information bits transmitted by IM,
in Section~\ref{S:Prec}, where it is also shown that FTN signaling 
can be seen as precoding.
In Section~\ref{S:MMSE}, we discuss low-complexity detectors
for SCIM signals over ISI channels.
To find an approximation
solution to an optimal detection for a better performance,
a detector based on variational inference \cite{Jordan99}
that is widely used in machine learning
is considered in
Section~\ref{S:VI}. In Section~\ref{S:G_SCIM},
SCIM with precoding is generalized to multiple access channel
so that multiple IoT device can share the same 
radio resource block for uplink transmissions.
Simulation results are presented in Section~\ref{S:Sim}.
The paper is concluded with some remarks in Section~\ref{S:Conc}.

{\it Notation}:
Matrices and vectors are denoted by upper- and lower-case
boldface letters, respectively.
The superscripts $\rT$ and $\rH$
denote the transpose and complex conjugate, respectively.
The $p$-norm of a vector $\ba$ is denoted by $|| \ba ||_p$
(If $p = 2$, the norm is denoted by $||\ba||$ without
the subscript).
The support of a vector is denoted by ${\rm supp} (\bx)$
(which is the index set of the non-zero elements of $\bx$).
The superscript $\dagger$ denotes the pseudo-inverse.
For a vector $\ba$, ${\rm diag}(\ba)$
is the diagonal matrix with the diagonal elements
from $\ba$. For a matrix $\bX$
(a vector $\ba$), $[\bX]_n$ ($[\ba]_n$) represents
the $n$th column (element, resp.).
If $n$ is a set of indices,
$[\bX]_n$ is a submatrix of $\bX$ obtained by
taking the corresponding columns.
$\uE[\cdot]$ and ${\rm Var}(\cdot)$
denote the statistical expectation and variance, respectively.
$\cC \cN(\ba, \bR)$
($\cN(\ba, \bR)$)
represents the distribution of
circularly symmetric complex Gaussian (CSCG)
(resp., real-valued Gaussian)
random vectors with mean vector $\ba$ and
covariance matrix $\bR$.

\section{Single-Carrier Index Modulation} \label{S:SM}

In this section, we present SCIM
that is introduced in \cite{JC_ICC17} \cite{Nakao17}.

\subsection{System Model}

We consider SC transmission over 
an ISI channel with cyclic prefix (CP) \cite{Falconer02}
from a device to a base station (BS) or access point (AP).
Let $\bd = [d_0 \ \ldots \ d_{L-1}]^\rT$
denote a block of data symbols, denoted by $\{d_l\}$,
to be transmitted over an ISI
channel, where $L$ is the length of $\bd$.
Then, the received signal at time $l$ is given by 
\begin{align}
r_l = \sum_{p=0}^{P-1} h_p d_{l-p} + n_l, 
\end{align}
where $h_p$ is the $p$th coefficient of the ISI
channel of length $P$ and $n_l \sim \cC \cN(0, N_0)$
is the background noise.
For the transmission of each
block without inter-block 
interference (IBI), a CP is appended to $\bd$. At the BS
after removing the signal corresponding to CP, we have
\begin{align}
\br & = [r_0 \ \ldots \ r_{L-1}]^\rT \cr
& = \cH \bd + \bn,
	\label{EQ:br}
\end{align}
where 
$\bn = [n_0 \ \ldots \ n_{L-1}]^\rT$ and
$\cH$ is a circulant matrix that is given by
\begin{align*}
\cH & = \left[
\begin{array}{llll}
h_0 & h_{L-1} & \cdots & h_{1} \cr
h_1 & h_{0} & \cdots & h_{2} \cr
\vdots & \vdots & \ddots & \vdots \cr
h_{L-1} & h_{L-2} & \cdots & h_{0} \cr
\end{array}
\right].
\end{align*}
Here, $h_{P} = \ldots = h_{L-1} = 0$ for $L > P$.

Unlike conventional SC,
$\bd$ in SCIM is $Q$-sparse, i.e., $\bd \in \Sigma_Q$,
where
$$
\Sigma_Q = \{\bd \,\bigl|\, ||\bd||_0 \le Q\}.
$$
Throughout the paper, 
we assume that the sparsity of $\bd$ is $Q$
(i.e., there are $Q$ non-zero elements in $\bd$)
and the non-zero symbols are referred to as
active symbols. In addition,
$\bd$ is referred to as an SCIM symbol and $L$
is equivalent to as the slot length. That is,
one SCIM symbol is to be transmitted within a slot.
In addition, we assume that a non-zero element of $\bd$
(i.e., an active symbol) is an element of an $M$-ary
constellation, i.e., $d_l \in \cS$ if $d_l \ne 0$,
where $\cS$ is the signal constellation and 
$|\cS|=M$. 
We also assume that zero is not
an element of $\cS$, i.e., $0 \notin \cS$
and a non-zero element of $\bd$, i.e., $d_l \in \cS$, has
the following properties:
\begin{align*}
\uE[d_l] = 0 \ \mbox{and} \ {\rm Var}(d_l) = \sigma_d^2,
\end{align*}
where $\sigma_d^2$ represents the (active) symbol energy.
For example, if we consider binary phase shift keying (BPSK)
for $\cS$ with $\cS = \{-A, A\}$, we have $\sigma_d^2 = A^2$.
Then, the number of information bits per slot becomes
\be
N_{\rm b}(L,Q,M) = \lfloor \log_2 \binom{L}{Q} \rfloor + Q \log_2 M.
	\label{EQ:Nb}
\ee
The resulting system is referred to as SCIM in this paper.
SCIM can be seen as a time-domain version of 
IM with single cluster in \cite{Basar13} or a generalization
of pulse-position modulation (PPM). To see that PPM
is a special case of SCIM, we can assume that
$d_l \in \{A, 0\}$ and $Q = 1$, which becomes a $Q$-ary PPM.

As mentioned earlier, since SCIM does not require 
IFFT and has a low PAPR as an SC transmission scheme \cite{Falconer02},
it might be attractive for IoT devices of limited complexity.

Note that the number of information bits transmitted by IM
in \eqref{EQ:Nb} can be maximized if $Q = \frac{L}{2}$ for an even $L$.
However, large $Q$'s not only degrade the energy
efficiency, but also increase the complexity
of the signal detection including the ML detection
(as will be explained in Subsection~\ref{SS:OD}).
To avoid the high computational complexity,
multiple clusters can be considered as in OFDM-IM \cite{Basar13},
where the ML detection is independently carried out for each cluster.
However, clusters are not orthogonal in SCIM
due to ISI channels unless $P = 1$ (i.e., $\cH$ is diagonal).
Therefore, unlike OFDM-IM, the use of multiple clusters 
does not help reduce the computational complexity in SCIM.

\subsection{Bit-to-Index Mapping}

For a large $L$, the number of information bits transmitted by 
IM, $N_{\rm im} = \lfloor \log_2 \binom{L}{Q} \rfloor$ is also large 
and a non-trivial bit-to-index mapping rule exists.
For example, if $L = 64$ and $Q = 4$, there are $19$ bits
that can be transmitted by IM and a certain mapping rule
from 19 bits to $2^{19} = 524,288$ active index sets can be used.
At the BS, a demapping rule has to be used to
decide 19 bits from the estimated index set of active symbols.
A look-up table approach can be used for a demapping rule.
However, it may require a large memory.
To avoid this difficulty, 
we can impose a certain structure for IM.

Suppose that the block can be divided into $Q$ subblocks 
(or clusters) and
each subblock consists of $D$ symbols,
where $L = Q D$. Here, $D = L/Q$ is assumed to be a positive integer.
It is assumed that only one symbol per subblock is active 
and there are $Q$ active symbols per block as before.
For convenience,
the resulting IM, which can be seen as $D$-ary PPM for each subblock,
is referred to the structured IM (with $Q$ active symbols) in this
paper. Clearly, 
in this case, we only need a mapping table for $D$-ary PPM for IM.

In the structured IM,
the index set of active 
symbols or the support of $\bs$ has the following constraint:
\be
{\rm supp}(\bs) \in \cL_Q = \left\{\{l_0, \ldots,l_{Q-1}\}\,|\, 
l_q \in [qD, (q+1)D], \forall  q \right\},
\ee
where $\cL_Q$ represents the set of all the possible
supports of $\bs$ of the structured IM with $Q$ active symbols.
The number of bits transmitted per block in the structured IM,
which is denoted by $\tilde N_{\rm im}$,
becomes
\be
\tilde N_{\rm im} = Q \lfloor \log_2 \frac{L}{Q} \rfloor
 = Q \lfloor \log_2 D \rfloor.
	\label{EQ:tN}
\ee
If $L = 64$ and $Q = 4$, we have
$\tilde N_{\rm im} = 16 < N_{\rm im} = 19$.
In general, we have $\tilde N_{\rm im} < N_{\rm im}$.
However, when $L$ is sufficiently large,
we can show that $\tilde N_{\rm im}$ can approach $N_{\rm im}$ under
certain conditions as follows.

\begin{mylemma}	\label{L:1}
Suppose that $D$ is a power of 2 and $1/D \ll 1$.
With a fixed $D$, we have
\be
\lim_{L \to \infty}
\frac{\tilde N_{\rm im}}
{N_{\rm im}} \to 1, \ L \to \infty.
	\label{EQ:L1}
\ee
\end{mylemma}
\begin{IEEEproof}
When $D$ is fixed, we have $Q = O(L)$. Then,
from \cite{Spencer14}, it can be shown that
\begin{align}
N_{\rm im} = \lfloor \log_2 \binom{L}{Q} \rfloor 
=\lfloor (1+ o(1)) L H_{\rm b}(Q/L) \rfloor,
\end{align}
where $H_{\rm b} (p) = -p \log_2 p - (1-p) \log_2 (1-p)$
is the binary entropy function.
Since $Q/L = 1/D\ll 1$, we have
$$
H_{\rm b}(Q/L) \approx \frac{Q}{L} \log_2 \frac{L}{Q}.
$$
Thus, it follows
\begin{align*}
N_{\rm im} \approx \lfloor Q \log_2 \frac{L}{Q} \rfloor
= Q \log_2 D,
\end{align*}
since $D$ is assumed to be a power of 2.
From \eqref{EQ:tN}, we can also show that $\tilde N_{\rm im}
= Q \log_2 D$. Thus, as $L \to \infty$, we have
\eqref{EQ:L1}, which completes the proof.
\end{IEEEproof}

While it is desirable to have a sufficiently
small number of active symbols, i.e., $Q \ll L$,
for a high energy efficiency, 
it is also highly desirable that $D$ is a power of 2 in the structured IM
according to Lemma~\ref{L:1}.

\section{SCIM with Precoding}	\label{S:Prec}

The number of information bits that can be transmitted by IM,
i.e., $N_{\rm im} = \lfloor \log_2 \binom{L}{Q} \rfloor$,
depends on the length of block, $L$.
If $Q$ is fixed due to an energy constraint,
we need to increase $L$ for a larger $N_{\rm im}$,
which however results in the increase of the system bandwidth
(for a fixed symbol interval).
Without increasing the block length, $L$, it might be possible to increase
the number of information bits using precoding.
In this section, we generalize SCIM with precoding 
\cite{JC_ICC17} and discuss
its relation to FTN signaling \cite{Mazo75}.

\subsection{Precoding}

Let $\bPsi$ be a precoding matrix of size
$L \times N$, where $N \ge L$. 
Denote by $\bpsi_n$ the $n$th column of $\bPsi$,
i.e., $\bPsi = [\bpsi_0 \ \ldots \ \bpsi_{N-1}]$.
Then, the precoded SCIM symbol is given by
$$
\bd = \bPsi \bs =\sum_{n=0}^{N-1} \bpsi_n s_n.
$$
Here, $\bd$ is no longer sparse, but it has a sparse
representation where $\bs = [s_0 \ \ldots \ s_{N-1}]^\rT$ is sparse.
Clearly, 
the length of SCIM symbol becomes $N$, not $L$, and
more bits can be transmitted by IM  as $\binom{N}{Q} > \binom{L}{Q}$
for $N > L$. 
In SCIM with precoding, $\bPsi \bs$ is transmitted with CP.
Thus, the received signal at the BS after removing CP becomes 
\begin{align}
\br & = \cH \left( \sum_{n=0}^{N-1} \bpsi_n s_n \right) + \bn \cr
& = \cH \bPsi \bs + \bn.
	\label{EQ:br2}
\end{align}

Define the discrete Fourier transform (DFT) 
matrix as
$$
[\bF]_{m,l} = \frac{1}{\sqrt{L}}
e^{- \frac{j 2 \pi ml}{L}}, \ m, l = 0, \ldots, L-1.
$$
If $\bPsi = \bF^\rH$, SCIM with precoding becomes OFDM-IM.
To see this, we apply the DFT to $\br$. Then, 
from \cite{ChoiJBook}, we have
\begin{align*}
\by & = \bF \br \cr
& = \bF \cH \bF^\rH \bs + \bF \bn \cr
& = \bH \bs + \tilde \bn,
\end{align*}
where $\tilde \bn = \bF \bn$ and $\bH$ 
is a diagonal matrix, which is referred
to as the frequency-domain channel matrix and given by
$$
\bH = {\rm diag} (H_0, \ldots, H_{L-1}).
$$
Here, 
$H_l = \sum_{p=0}^{P-1} h_p e^{- \frac{j 2 \pi p l}{L}}$.
As mentioned earlier, since OFDM-IM has a poor PAPR performance,
$\bPsi = \bF^\rH$ is not desirable as a precoding matrix, and
a different precoding
matrix is to be chosen to avoid a high PAPR. 
To this end, it is desirable to have $\bpsi_n$ that has
a high energy concentration at a certain time. Note that
if $\bpsi_n = \bu_n$, where the $\bu_n$'s 
represent the standard basis vectors,
a good PAPR performance is achieved. However, 
there is no gain in terms of $N_{\rm im}$,
because the resulting precoding matrix is $\bPsi = \bI$.
To find good precoding matrices,
there might be a number of different approaches 
under various constraints. 
For example, in \cite{JC_ICC17},
a design approach to achieve 
repetition diversity gain with precoding is considered.

\subsection{FTN Signaling}

In \cite{Ishihara17}, the notion of 
FTN signaling \cite{Mazo75}
is applied to SCIM.
In FTN signaling, the symbol transmission rate
can be higher than the Nyquist rate (of a given bandwidth)
by a factor of the inverse of the time-squeezing factor 
\cite{Fan17}.
SCIM with FTN signaling 
can be seen as an example of SCIM 
with precoding where the precoding matrix depends on
the time-squeezing factor, denoted by $\xi\ (\le 1)$, and shaping pulse
(which is the impulse response of the transmit 
filter).
With FTN signaling,
the size of precoding matrix becomes $L \times \lfloor \frac{L}{\xi} \rfloor$,
i.e., $N = \lfloor \frac{L}{\xi} \rfloor$,
and $\bPsi$ has its coefficients that are decided by
the time-squeezing factor and shaping pulse.
For example, if the Nyquist pulse is used, i.e.,
$$
g(t) = \frac{\sin \pi \frac{t}{T}}{ \pi \frac{t}{T} },
$$
where $T$ is the Nyquist sampling interval,
the $(l,n)$th element of $\bPsi$
can be given by
\be
\psi_{l,n} = [\bPsi]_{l,n}  = g( (l - n \xi)T),
	\label{EQ:psi_g}
\ee
where $l \in \{0, \ldots, L-1\}$ and
$n \in \{0, \ldots, N-1\}$.
The resulting SCIM with the precoding matrix in
\eqref{EQ:psi_g} is referred to as 
SCIM with FTN precoding in this paper.
Note that it is also called FTN-IM in \cite{Ishihara17}. 

Due to the time-squeezing factor $\xi$ in FTN signaling,
there might be severe ISI, which can be overcome by
equalizers and coding \cite{Mazo75} \cite{Anderson13}
with the sampled signals at a sampling rate of $\frac{1}{\xi T}$.
To detect active symbols
in SCIM with FTN precoding, 
a higher sampling rate 
(i.e., $\frac{1}{\xi T}$)
can be used as in \cite{Ishihara17}. 
However, it is also possible to detect
active symbols with the Nyquist sampling rate
(or a lower rate than the Nyquist sampling rate)
by exploiting the sparsity of $\bs$,
which will be discussed in Subsection~\ref{SS:CS}.

\section{MMSE and CS-based Detection}	\label{S:MMSE}

In this section, we discuss low-complexity detection methods
for SCIM without and with precoding.

\subsection{MMSE Detection}	\label{SS:MMSE}

In this subsection, we assume that $\bPsi = \bI$,
i.e., no precoding is employed for SCIM.
As in \cite{Falconer02, Pancaldi08}, the frequency domain 
equalization (FDE) can be considered to detect $\bs$
(regardless of its sparsity)
with low-complexity.
To this end, we can apply 
the DFT to $\br$ in \eqref{EQ:br}, and it can be shown that
\begin{align}
\by & = \bF \br \cr
& = \bH \bF \bs + \tilde \bn ,
	\label{EQ:by}
\end{align}

In FDE, we estimate $\bx = \bF \bs$ 
(instead of $\bs$) using
the minimum mean squared error (MMSE) filter 
(which is a single-tap equalizer)
that is given by
\begin{align}
\bW_{\rm mmse} & = \uE[\bx \by^\rH] \left(
\uE[\by \by^\rH] \right)^{-1} \cr
& = \bH^\rH \left( 
\bH \bH^\rH + \frac{L}{Q \gamma} \bI \right)^{-1} \cr
& = {\rm diag}\left(
\frac{H_0^*}{|H_0|^2 + \frac{L}{Q \gamma} }, \ldots,
\frac{H_{L-1}^*}{|H_{L-1}|^2 + \frac{L}{Q \gamma} }
\right),
	\label{EQ:Wmmse}
\end{align}
where
$\gamma = \frac{\sigma_s^2}{N_0}$.
Here, $\sigma_s^2$ represents the variance of non-zero $s_l$.
Note that in \eqref{EQ:Wmmse}, if we assume that
the active symbols of $\bs$ are uniformly distributed,
we have
$$
\uE[\bx \bx^\rH] 
= \uE[\bs \bs^\rH] 
= \frac{Q  \sigma_s^2}{L} \bI.
$$
Once $\bx$ is estimated
as $\bW_{\rm mmse} \by$, $\bs$ can be recovered by taking 
inverse DFT (IDFT).
That is,
\begin{align}
\hat \bs_{\rm mmse} 
& = \bF^{-1} \bW_{\rm mmse} \by \cr
& = \bF^\rH \bW_{\rm mmse} \by.
	\label{EQ:smmse}
\end{align}
From the estimate of $\bs$ in 
\eqref{EQ:smmse},
the largest $Q$ elements in terms of their amplitudes
can be chosen for the detection of index modulated signals.
Throughout the paper, the resulting detector is referred to
as the MMSE detector.

\subsection{CS-based Detection}	\label{SS:CS}


In SCIM with FTN precoding,
if the receiver chooses the Nyquist sampling rate,
the number of received signals, $L$,
becomes smaller than the block length, $N$.
In general, when precoding with $N > L$ is used,
the resulting system in \eqref{EQ:br2}
becomes overdetermined, while $\bs$ is sparse.
Thus, the notion of CS 
\cite{Donoho06} \cite{Candes06}
can be exploited to estimate $\bs$ from $\br$.
For example, as in \cite{Choi_Ko15} \cite{JC_ICC17},
a CS-based detector can be considered using the sparsity of $\bs$.

Since $\bs$ is sparse, 
the estimation of $\bs$ can be carried out via 
$\ell_1$-minimization as follows:
\be
\hat \bs = \argmin ||\bs||_1 \
 \mbox{subject to} \ ||\br-\cH \bPsi \bs||^2 \le \epsilon,
	\label{EQ:CSF}
\ee
where $\epsilon$ is the error bound.
The formulation in \eqref{EQ:CSF} is a typical CS problem
\cite{Candes08}
and a number of algorithms are available to obtain $\hat \bs$
\cite{Eldar12}.

In \eqref{EQ:CSF},
the recovery guarantee depends on 
the restricted isometry property (RIP) of $\cH \bPsi$
\cite{Candes08b} \cite{Baraniuk08}.
It is said that $\bA$ satisfies the RIP of order $k$
with RIP constant $\delta_k \in (0,1)$ if
there exists a $\delta_k$ such that
$$
(1- \delta_k) ||\bx||^2 \le ||\bA \bx||^2 \le (1+\delta_k) ||\bx||^2,
$$
where $\bx \in \Sigma_k$. If $\bA$ has unit-norm columns,
the RIP constant is also related to the coherence of $\bA$ 
as follows \cite{Eldar12}:
\be
\delta_k = (k-1) \mu(\bA),
	\label{EQ:dmu}
\ee
where $\mu(\bA)$ is the coherence of $\bA$ that is defined as
$$
\mu(\bA) = \max_{1 \le n < m \le N}
\frac{|\ba_n^\rH \ba_{m}|}{||\ba_n|| \cdot ||\ba_{m}||},
$$
where $N$ is the number of the columns of $\bA$.

Provided that $\cH \bPsi$ satisfies the RIP,
it is known that if 
\be
L \ge C Q \ln \frac{N}{Q},
	\label{EQ:LQN}
\ee
where $C$ is a constant that is independent of $N$ and $Q$,
$\bs$ can be recovered by solving \eqref{EQ:CSF} with a high probability
\cite{Candes08} \cite{Baraniuk08} \cite{Eldar12}.

If FTN precoding is employed,
we have $L = \xi N$.
Suppose that $\tau = \frac{N}{Q}$ is fixed when $N$ increases.
Then, from \eqref{EQ:LQN}, we have
$$
L = C Q \ln \tau \ \mbox{or} \ \xi = C \frac{\ln \tau}{\tau},
$$
which implies that
the time-squeezing factor $\xi$ can be quite small
for a large $\tau$. 
For example, if $C = 0.28$ \cite{Eldar12} and $\tau = 100$,
we have $\xi = 0.012$.
Thus, we can expect to be able to greatly increase 
the number of information bits transmitted by IM,
$N_{\rm im}$, using FTN precoding.
However, in practice, 
$\cH \bPsi$ may not satisfy
the RIP as $\cH$ is decided by a random ISI channel.
In addition, the error probability of
successful recovery is often too high 
to meet the requirement in wireless communications, say $10^{-3}$ or lower.
Consequently, $\xi$ cannot be too small.

Although \eqref{EQ:CSF} is a convex optimization problem,
its computational complexity can be high. Thus,
low-complexity greedy algorithms might be used to find
approximate solutions to \eqref{EQ:CSF}.
For example, the 
(orthogonal matching pursuit)
OMP algorithm \cite{Pati93} \cite{Davis94} can be used
as a low-complexity approach to estimate sparse $\bs$.
In general, the computational complexity of the OMP algorithm
depends on the size of the measurement matrix and
sparsity. Provided that $Q$ is sufficiently small,
the computational complexity becomes $O(LN)$ \cite{Eldar12}.

\subsection{CS-based Detection with MMSE Filtering}	\label{SS:CS_MMSE}

\begin{figure}[thb]
\begin{center}
\includegraphics[width=\figwidth]{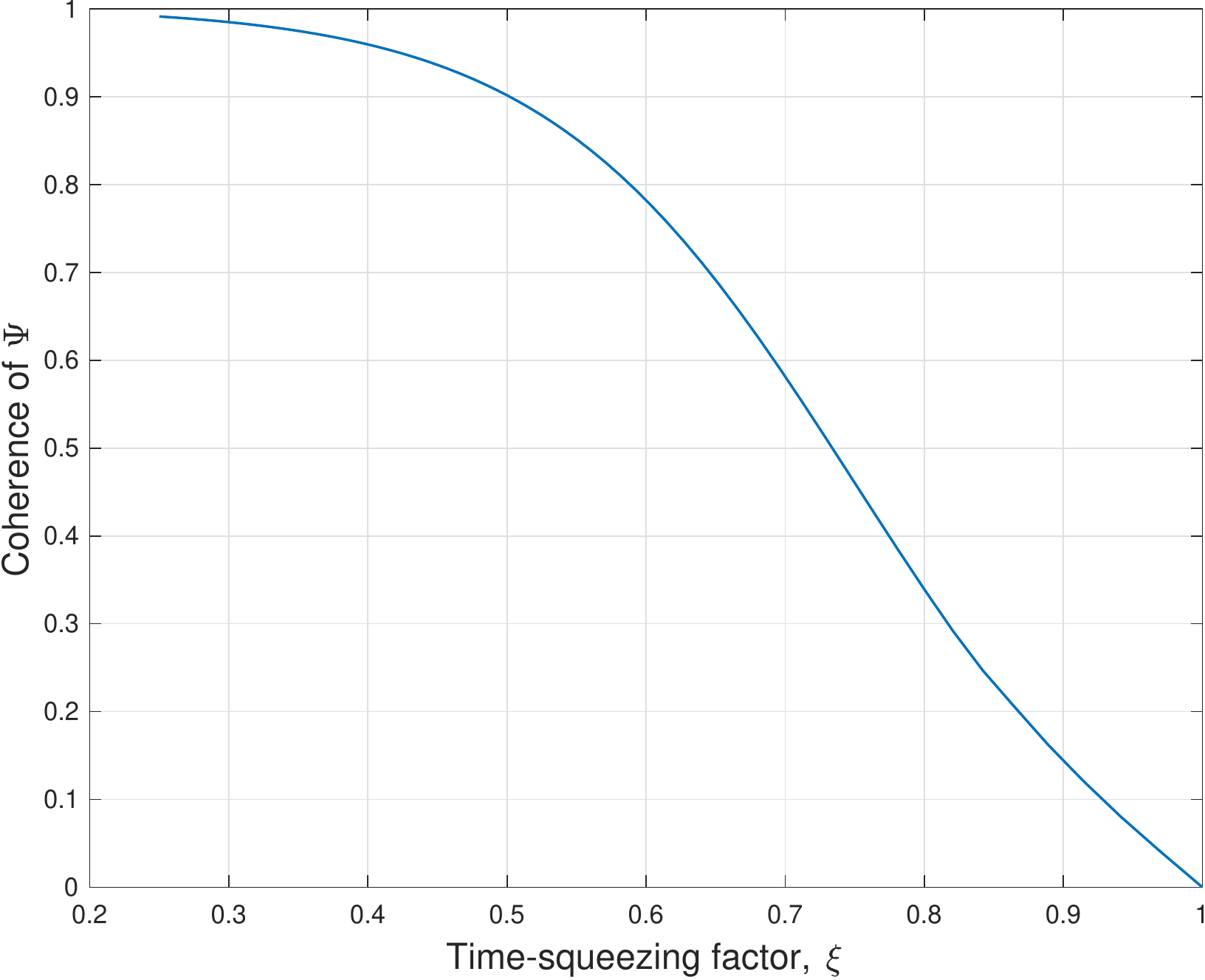} 
\end{center}
\caption{The coherence of $\bPsi$ in FTN precoding with
$L = 64$ as a function of the time-squeezing factor, $\xi$.}
        \label{Fig:plt_coh}
\end{figure}

Without precoding,
the MMSE filter is able to provide an estimate of $\bF \bs$
from $\by$ as in \eqref{EQ:smmse}.
We can also use the MMSE filter with precoding to 
estimate $\bF \bPsi \bs$ (instead of $\bF \bs$).
Since $\bF$ is unitary,
we can find an estimate of $\bPsi \bs$ as follows:
\be
\bv = 
\bF^\rH \bW_{\rm mmse} \by
\approx \bPsi \bs.
\ee
Then, from $\bv$, 
the following optimization problem can be considered to estimate
$\bs$:
\be
\hat \bs = \argmin ||\bs||_1 \
 \mbox{subject to} \ || \bv - \bPsi \bs||^2 \le \epsilon.
	\label{EQ:CSF_v}
\ee
Provided that $\bv$ is a good estimate of $\bPsi \bs$,
the performance of recovering $\bs$ from $\bv$ depends on $\bPsi$.
According to \eqref{EQ:dmu},
for a certain RIP constant, it is required that
$Q$ is to be inversely proportional 
the coherence of $\bPsi$ 
(i.e., for a large $Q$, the coherence of $\bPsi$ has to be small).
We show the coherence of $\bPsi$ in FTN precoding
with $L = 64$ in Fig.~\ref{Fig:plt_coh}.
Since the coherence increases as $\xi$ decreases,
as $N$ increases (i.e., $\xi$ decreases) for a fixed $L$, $Q$ has to be small.

For a low-complexity approximation, 
we can again use the OMP algorithm to 
recover $\bs$ from $\bv$. That is, to solve \eqref{EQ:CSF_v},
the OMP algorithm can be used.
The resulting approach can provide a good performance
if $\bPsi$ satisfies the RIP
and the MMSE filtering provides a good estimate of $\bPsi \bs$.

\section{Variational Inference based Detection}	\label{S:VI}

In this section, we study a different approach
based on variational inference 
\cite{Jordan99} \cite{Bishop06}
to detect the indices of active symbols.
This approach does not solve \eqref{EQ:CSF}, but
finds an approximation of the maximum
a posteriori probability (MAP) detection.

\subsection{Optimal Detection}	\label{SS:OD}

In order to estimate $\bs$, we can consider the 
ML approach.
From \eqref{EQ:br2},
the likelihood function of $\bs$ is given by
$$
f(\br\,|\, \bs) = 
\frac{1}{(\pi N_0)^L}
\exp\left( - \frac{1}{N_0}||\br - \cH \bPsi \bs||^2 \right).
$$
The ML estimate can be found as \cite{ChoiJBook2}
\begin{align}
\hat \bs 
& = \argmax_{\bs \in \bar \Sigma_Q} f(\br\,|\, \bs) \cr
& = \argmin_{\bs \in \bar \Sigma_Q} ||\br - \cH \bPsi \bs||^2,
	\label{EQ:MLD}
\end{align}
where
$\bar \Sigma_Q = \{\bs \, |\, \bs \in \Sigma_Q,
\ s_l \in \bar \cS \}$.
Here, $\bar \cS = \cS \cup \{0\}$.
We note that 
$|\bar \Sigma_Q| = 2^{N_{\rm b}}$, which grows exponentially with $Q$
(from \eqref{EQ:Nb}).
Since the complexity of the ML detection
is proportional to $|\bar \Sigma_Q|$,
it might be computationally prohibitive unless $Q$ is small
(i.e., $Q$ is 1 or 2)
if an exhaustive search is used. 

In order to take into account the sparsity of $\bs$,
we can consider the MAP detection.
Let $\Pr(\bs \,|\, \br)$ represent the a posteriori probability
of $\bs$ for given $\br$.
Since
\be
\Pr(\bs\,|\,\br) \propto f(\br\,|\, \bs) \Pr(\bs),
\ee
where $\Pr(\bs)$ is the a priori probability of $\bs$
(which can take into account the sparsity of $\bs$),
the MAP detection to find $\bs$ 
that maximizes $\Pr(\bs \,|\, \br)$ can be given by
\be
\hat \bs 
= \argmax_{\bs \in \bar \Sigma_Q} \Pr(\bs \,|\, \br).
\ee
Like the ML detection, if an exhaustive search is considered to perform
the MAP detection, the complexity becomes prohibitively high.

\subsection{Variational Inference}

Since the optimal detection (ML or MAP)
requires a high computational complexity,
we may need to consider a low-complexity
approach to find an approximation.
To this end, we can consider the variational inference
\cite{Jordan99} \cite{Bishop06} \cite{Blei17}, which is
a well-known machine learning technique.

The variation inference is to obtain an approximate
solution to the MAP problem using variational distributions of
$\bs$.
Let $\rho_n(s_n)$ be the distribution of $s_n \in \bar \cS$.
In addition, $\cR$ denotes the set of the distributions of $\rho_n
(s)$ for $s \in \bar \cS$.
Furthermore, we assume that the $s_n$'s are independent. 
Thus, we have
$$
\rho(\bs) = \prod_{n=0}^{N-1} \rho_n (s_n),
$$
which results in the mean-field approximation in variational inference
\cite{Bishop06}.
An approximation of the a posteriori probability,
$\Pr(\bs\,|\, \br)$, can be obtained through the following optimization:
\be
\rho^*  = \argmin_{ \rho  \in \cR^N}
{\rm KL} \left(\rho (\bs) || \Pr( \bs\,|\,\br) \right),
	\label{EQ:KL}
\ee
where ${\rm KL}(\cdot)$ 
is the Kullback-Leibler (KL) divergence \cite{CoverBook},
which is defined as
$$
{\rm KL} (\rho(\bs)||f(\bs))
= \sum_\bs \rho(\bs) \ln \frac{\rho(\bs)}{f(\bs)}.
$$
Here, $f(\bs)$ is any distribution of $\bs$
with $f(\bs) > 0$ for all $\bs \in \bar \cS^N$.
Since the KL divergence is 
to measure the difference between two probability distributions,
$\rho^*(\bs)$ in \eqref{EQ:KL} becomes an approximation
of $\Pr(\bs\,|\, \br)$.
Then, the MAP detection can be carried out 
with $\rho^*(\bs)$ instead of $\Pr(\bs\,|\, \br)$.
Thanks to the assumption that
the $s_n$'s are independent, 
$s_n$ can be estimated as follows:
$$
\hat s_n = \argmax_{s_n \in \bar \cS} \rho_n^* (s_n).
$$
If $\hat s_n \ne 0$, $s_n$ is detected as an active symbol
and its value in the $M$-ary
constellation, $\cS$, can also be obtained.

Note that since the sparsity of $\bs$ is $Q$,
it has to be taken into account. 
To this end, let
$\tilde \rho_n = 1 - \rho_n (0)$.
In addition, denote by 
$a(n)$ the index of the $n$th largest one among
$\{ \tilde \rho_0, \ldots, \tilde \rho_{N-1}\}$,
i.e.,
$\tilde \rho_{a(0)} \ge \ldots \ge \tilde \rho_{a(N-1)}$.
Then, the index set of $Q$ active symbols becomes
$\{a(0), \ldots, a(Q-1)\}$.
We can also readily impose the constraint of structured IM if
$\bs$ is a structured IM signal in choosing $Q$ active symbols.
Furthermore, soft-decisions
on the IM bits can be found as $\rho(\cdot)$
is a probability, which might be useful for 
decoding when a channel code is used.

\subsection{CAVI Algorithm with Gaussian Approximation}

As shown in \cite{Blei17},
the minimization of the KL divergence in \eqref{EQ:KL}
is equivalent to the maximization of the evidence lower bound (ELBO),
which is given by
\be
{\rm ELBO} (\rho) = \uE[\ln f(\br, \bs)] - \uE[\ln \Pr(\bs)],
\ee
where the expectation is carried out over $\bs$.
Let $\bs_{-n} = [s_1 \ \ldots \ s_{n-1} \ s_{n+1} \ \ldots \ s_N]^\rT$.
Then, for given $\bs_{-n}$,
it can be shown that
\be
\rho_n \propto 
\exp \left( \uE_{-n} [\ln f(s_n\,|\, \bs_{-l}, \br) ] \right),
	\label{EQ:qE}
\ee
where the expectation,
denoted by $\uE_{-n}$, is carried out over $\bs_{-n}$.
The coordinate ascent variational inference (CAVI) algorithm
\cite{Bishop06, Blei17} is to update $\rho_n$, $n = 0,\ldots, N-1$,
while the other variational distributions, $\rho_{-n}$, are fixed.
The CAVI algorithm requires a number of iterations, denoted by $N_{\rm run}$.

To carry out the updating in \eqref{EQ:qE}, we need to have
a closed-form expression for 
$\uE_{-n} [\ln f(s_n\,|\, \bs_{-n}, \br) ]$.
Unfortunately, since $f(s_n\,|\, \bs_{-n}, \br)$ is a Gaussian mixture
and it is difficult to find a closed-form expression.
However, as shown in \cite{Choi_VI18},
a suboptimal approach is available with
the Gaussian approximation where
an active symbol $s_n$ is assumed to be a 
CSCG random variable.
In particular, let $s_l$ be a zero-mean CSCG random variable 
with variance $\sigma_s^2$ when $s_l \ne 0$.
Define the activity variable, $x_l$, as
$$
x_l = \left\{
\begin{array}{ll}
1, & \mbox{if $s_l \ne 0$} \cr
0, & \mbox{if $s_l = 0$.} \cr
\end{array}
\right.
$$
For convenience, let $\bA = \cH \bPsi$ and 
denote by $\ba_l$ the $l$th column of $\bA$.
Furthermore, let $\chi_l^{(i)} (x_l)$
be the estimate of $\Pr(x_l)$ in the $i$th iteration
of the CAVI algorithm,
where the superscript $(i)$ represents the $i$th iteration
(i.e., $i$ is used for the iteration index).
Under the Gaussian assumption of $s_l$,
the CAVI updating rule for $\chi_l^{(i)}(x_l)$ is given by
\begin{align}
\chi_l^{(i)} (x_l) & = e^{
- \br^\rH \bR_{l}^{(i)} (x_l)^{-1} \br
- \ln \det(\bR_l^{(i)} (x_l))}, \ x_l \in \{0,1\},
        \label{EQ:chil}
\end{align}
where
\begin{align}
\bR_l^{(i)} (x_l)
& =
\ba_l \ba_l^\rH x_l +
\sum_{t <l } \ba_t \ba_t^\rH \bar \chi_t^{(i)}  \cr
& \quad
+ 
\sum_{t >l } \ba_t \ba_t^\rH \bar \chi_t^{(i-1)} 
+ \gamma^{-1} \bI ,
        \label{EQ:iRl}
\end{align}
where $\bar \chi_l^{(i)}$ is the normalized version
of $\chi_l^{(i)}(x_l)$, which is given by
$\bar \chi_l^{(i)} =
\frac{\chi_l^{(i)}(1)}{\chi_l^{(i)}(0) + \chi_l^{(i)}(1)}$.
The resulting CAVI algorithm provides the support of $\bs$.
Once the support of $\bs$ is found, we can easily estimate the non-zero
elements of $\bs$ from $\br$.
In \cite{Choi_VI18}, detailed derivations
are presented and it is also shown that the complexity
per iteration is $O(L^2 N)$. From this, we can see that
the complexity grows linearly with $N$ 
(which is similar to the OMP algorithm) and
its complexity is higher than that of the OMP algorithm
by a factor of $L N_{\rm run}$.

\section{Generalization of SCIM to Multiple
Access for Low-Rate Devices}	\label{S:G_SCIM}

In this section, SCIM with precoding is generalized
to multiple access channel
so that multiple devices can be supported in the
same resource block.

Suppose that there are $K$ devices to transmit their signals to the BS.
Let $\cH_k$ and $\bPsi_k \in \uC^{L \times N}$ denote the channel
and precoding matrices of device $k$, respectively.
Then, the received signal at the BS is given by
\begin{align}
\br = \sum_{k=1}^K \cH_k \bPsi_k \bs_{k} + \bn.
\end{align}
Let 
\begin{align}
\bA &= [(\cH_1 \bPsi_1) \ \cdots \ (\cH_K \bPsi_K) ] \in \uC^{L \times
NK} \cr
\bz & = [\bs_1^\rT \ \ldots \ \bs_K^\rT]^\rT \in \uC^{NK \times 1}.
	\label{EQ:Az}
\end{align}
Then, it can be shown that
\be
\br = \bA \bz + \bn.
	\label{EQ:rAz}
\ee
If each device uses the structured IM with $Q$ active symbols
(and $N = DQ$),
$\bz$ can be seen as a structured IM with $QK$ active symbols.
It is clear that both the OMP and CAVI algorithms
can be employed at the BS to detect the 
signals from $K$ devices, i.e., $QK$-sparse
signal $\bz$, from $\br$.
For example, the CAVI algorithm can provide
an approximate solution to the MAP detection 
by obtaining an approximate a posteriori probability as follows:
$$
\rho (\bz) \approx \Pr(\bz\,|\br).
$$

Note that since the MMSE filtering is not applicable to the superposition
of precoded signals,
the approach in Subsection~\ref{SS:CS_MMSE} cannot be used.

\begin{figure}[thb]
\begin{center}
\includegraphics[width=\figwidth]{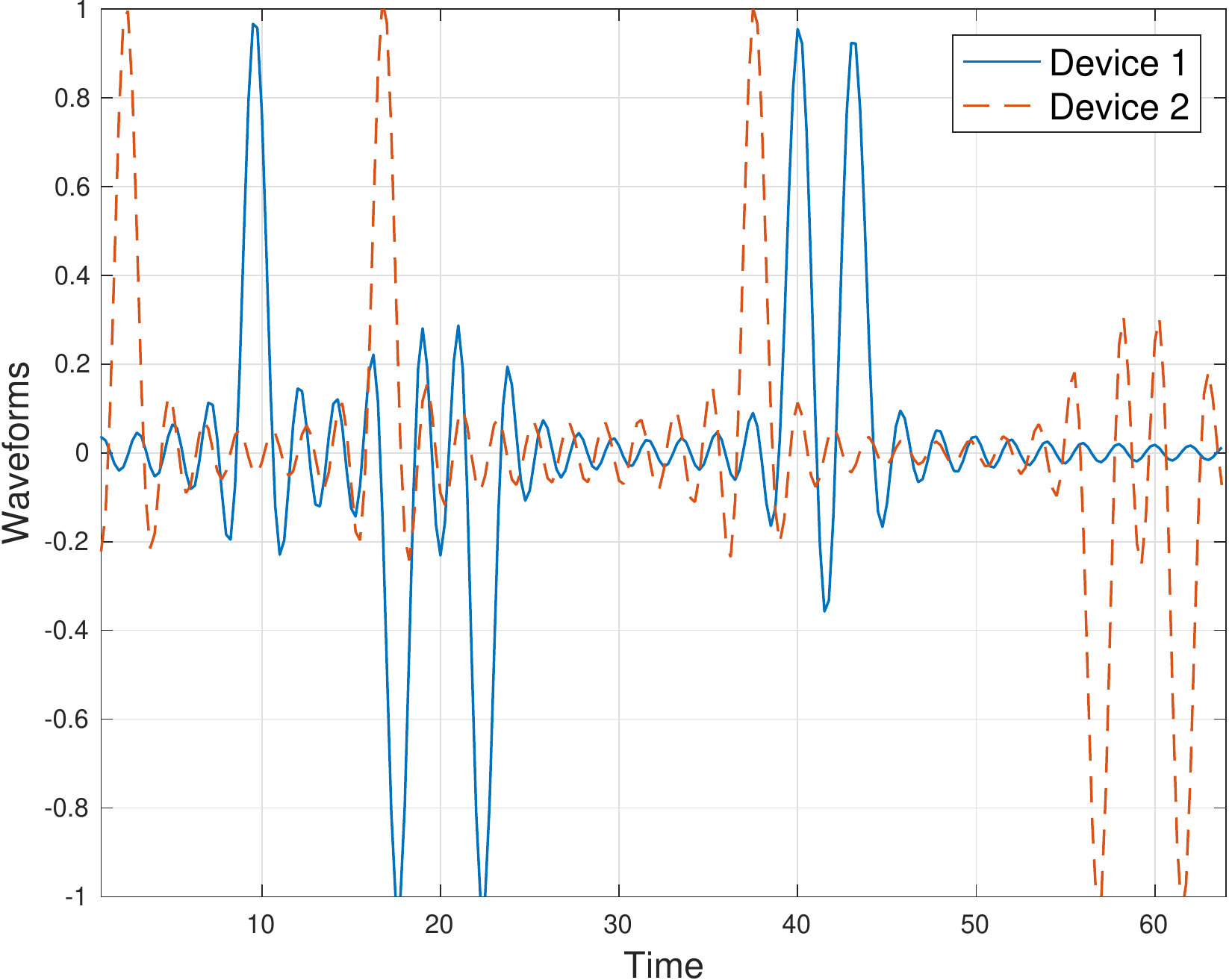} 
\end{center}
\caption{Transmitted SCIM waveforms 
transmitted from two low-rate devices
when $L = 64$ and $K = 2$ with $(Q, D) = (5, 8)$.}
        \label{Fig:plt_ma_ftn}
\end{figure}

FTN signaling can be applied to SCIM for multiple access.
In particular, according to \eqref{EQ:Az}, 
since the size of $\bA$ is $L \times NK$, 
the system time-squeezing factor becomes $\xi = \frac{L}{NK}$
with the following precoding matrix for device $k$:
\be
[\bPsi_k]_{l,n} = g((l- (nK + k-1) \xi)T), \ k = 1,\ldots,K, 
	\label{EQ:Pg2}
\ee
where $l \in \{0,\ldots, L-1\}$ and $n \in \{0,\ldots, N-1\}$.
Let $l_k = l - \xi (k-1)$. Then, from \eqref{EQ:Pg2},
we have
\be
[\bPsi_k]_{l,n} = g( (l_k- nK\xi)T),
	\label{EQ:PgK}
\ee
which shows that the effective time-squeezing
factor at a device becomes $K \xi = \frac{L}{N}$.
Thus, if $N < L$, the device's transmission rate
becomes lower than the BS's sampling rate
(or Nyquist rate). For example, if $L = 64$, $K = 2$,
and $(Q,D) = (5,8)$, we have $N = 40$.
Thus, the system time-squeezing factor 
is $\frac{64}{80} < 1$, while the device's time-squeezing
factor $K \xi$ is $\frac{64}{40} > 1$. 
In Fig.~\ref{Fig:plt_ma_ftn},
SCIM waveforms transmitted from 
two low-rate devices are illustrated
when $L = 64$ and $K = 2$ with $(Q, D) = (5, 8)$.
The BS would be able to recover the two sparse signals 
from $L = 64$ samples per block by solving \eqref{EQ:rAz}.

As shown above, 
the generalization of SCIM with FTN precoding 
to multiple access channel
can play a key role in supporting a number
of devices with a limited spectrum
by allowing multiple IoT
devices of low transmission rates to share the same
radio resource block. 
However, there are other issues to be addressed as follows.
\begin{itemize}
\item Channel Estimation: As shown in \eqref{EQ:Az},
the BS needs to know $\bA$, 
which means that it requires to estimate $\cH_k$
since $\bPsi_k$ is known.
As in \cite{3GPP_MTC} \cite{3GPP_NBIoT},
prior to data transmissions, a handshaking procedure can be used 
for random access where the channel estimation can also be carried out
as each active device is to transmit a preamble.
In particular, once the BS is able to detect the preambles
transmitted by multiple active devices without any collision,
it should be able to estimate their channels (taking the preambles
as pilot signals). Then, for data transmissions, 
multiple active devices can employ SCIM (with different
precoding matrices) to transmit their
signals in the same radio resource block.

\item When FTN precoding is used, in \eqref{EQ:PgK},
each device has the same effective time-squeezing factor, $\frac{L}{N}$.
In fact, it is also possible to have a different 
time-squeezing factor by allowing a different number 
columns of $\bPsi_k$, which is denoted by $N_k$. Then, the system
time-squeezing factor becomes $\frac{L}{\sum_{k=1}^K N_k}$,
while device $k$'s effective
time-squeezing factor is $\frac{L}{N_k}$.
In other words, supporting multiple devices of
different transmission rates is possible using SCIM with precoding.
This feature might be crucial to support a wide range of IoT devices
that may have different transmission rates.
\end{itemize}

\section{Simulation Results}	\label{S:Sim}

In this section, we present simulation results
of SCIM\footnote{In \cite{JC_ICC17} \cite{Nakao17},
performance comparisons between SCIM and OFDM-IM
can be found, where it is shown that SCIM outperforms OFDM-IM
in terms of error rates. Thus, we only present simulation
results of SCIM in this paper.} 
when 4-QAM (in this case,
$M = 4$) is used for active symbols
with $\cS = \{\pm 1 \pm j\}$ with the structured IM.
For the multipath channel, we assume that the channel coefficients
are independent and $h_p \sim \cC \cN(0, 1/P)$, 
$p = 0, \ldots, P-1$, i.e., a multipath Rayleigh fading channel
is considered.
The signal-to-noise ratio
(SNR) is defined as $\frac{E_{\rm b}}{N_0}$,
where $E_{\rm b}$ represents the bit energy that is given by
$$
E_{\rm b} = \frac{Q \sigma_s^2}{\tilde N_{\rm im} + Q \log_2 M}.
$$
Here, $\sigma_s^2 = 2$ as 4-QAM is used.
We also assume FTN precoding for SCIM.
For the signal detection, the following 3 different approaches
are considered at the receiver of BS: (i) the OMP 
detector that uses the OMP algorithm to solve \eqref{EQ:CSF};
(ii) the OMP-MMSE detector that employs
the MMSE filtering with the OMP algorithm to solve \eqref{EQ:CSF_v};
(iii) the CAVI detector that is based on the CAVI algorithm
(unless stated otherwise, we assume that the number of iterations
for the CAVI detector is set to $N_{\rm run} = 4$).
To see the performance, we consider the index error rate
(IER) that is the 
probability of erroneously
detection of any index of active symbols.

\begin{figure}[thb]
\begin{center}
\includegraphics[width=\figwidth]{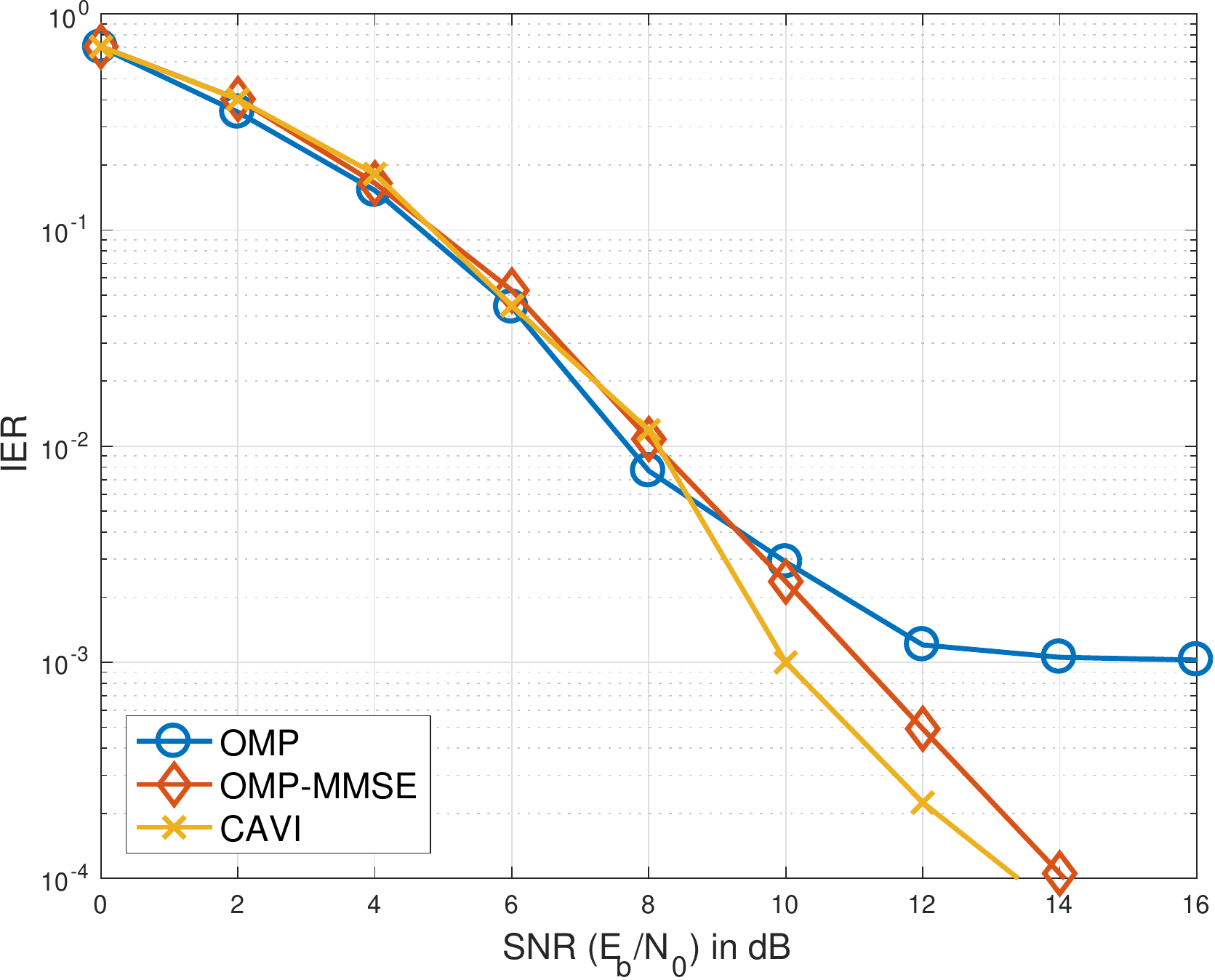} \\
(a) \\
\includegraphics[width=\figwidth]{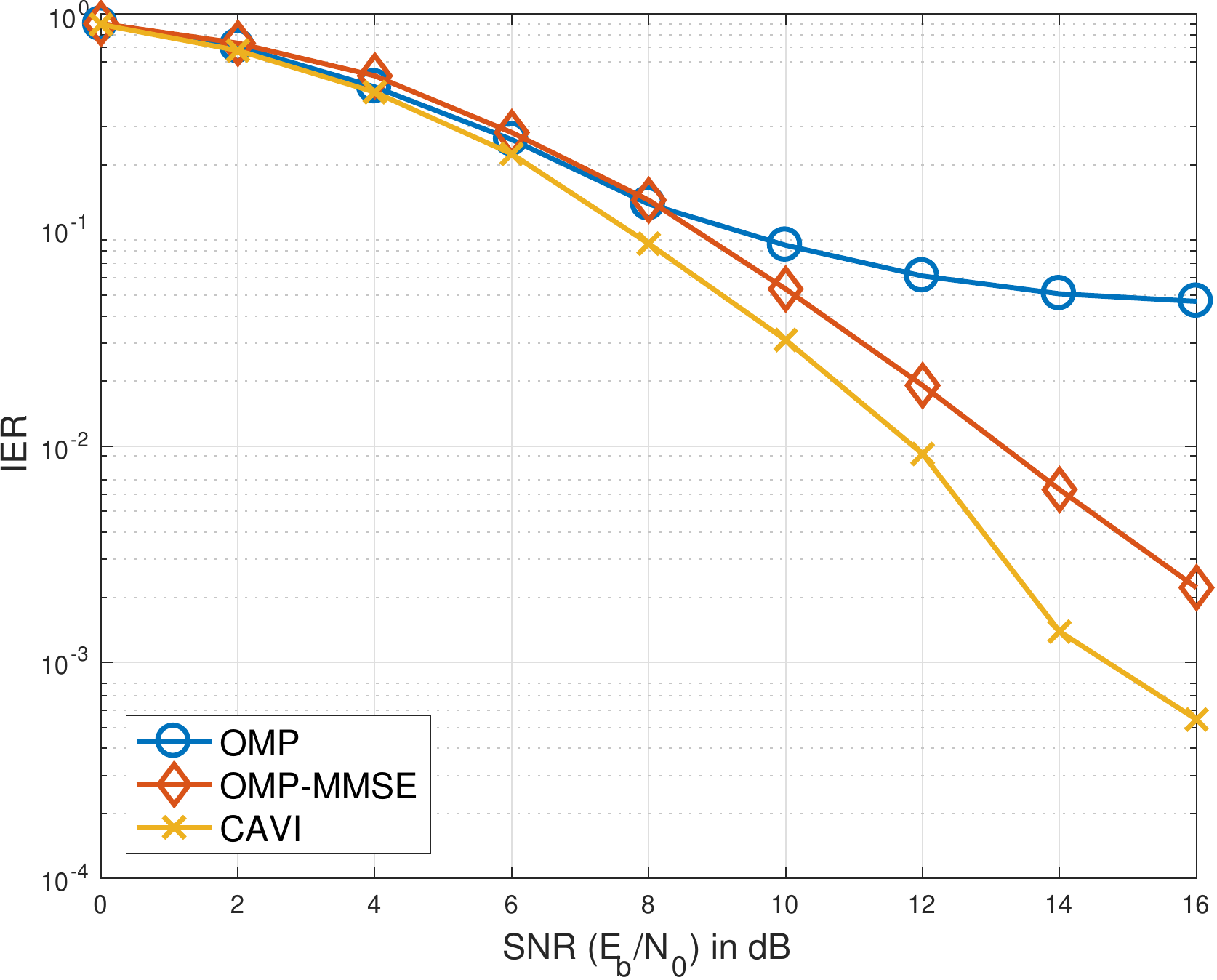} \\
(b) \\
\end{center}
\caption{IER as a function of SNR:
(a) $L = 128$, $P = 6$, $Q = 5$, $D = 32$;
(b) $L = 64$, $P = 4$, $Q = 10$, $D = 8$.}
        \label{Fig:plt12}
\end{figure}

We first consider SCIM with FTN precoding for a single device.
Fig.~\ref{Fig:plt12} shows the IERs of the three detectors
as functions of SNR with two different sets of parameters' values.
In Fig.~\ref{Fig:plt12} (a),
we consider $L = 128$ and $N = 160$
with $P = 6$ and $(D, Q) = (32, 5)$.
In this case, we have $\tilde N_{\rm im} = 25$ bits.
On the other hand, in Fig.~\ref{Fig:plt12} (b),
we consider $L = 64$ and $N = 80$ with $P = 4$ and $(D, Q) = (8, 10)$,
where $\tilde N_{\rm im} = 30$ bits.
Clearly, the size of the system in 
Fig.~\ref{Fig:plt12} (b)
is smaller than that 
in Fig.~\ref{Fig:plt12} (a), while the former transmits slightly more bits
than the latter. As a result, the performance of the system in 
Fig.~\ref{Fig:plt12} (a)
in terms of IER is better than that in Fig.~\ref{Fig:plt12} (b). 
We note that the CAVI detector
outperforms the other detectors, i.e., the OMP and OMP-MMSE
detectors, at the cost of a higher computational complexity.
It is also interesting to see that
the OMP-MMSE detector can provide a comparable performance
to the CAVI detector,
while the OMP detector suffers from the error floor.
Thus, we need to use the OMP-MMSE or CAVI detector at a high SNR
to avoid the error floor.

\begin{figure}[thb]
\begin{center}
\includegraphics[width=\figwidth]{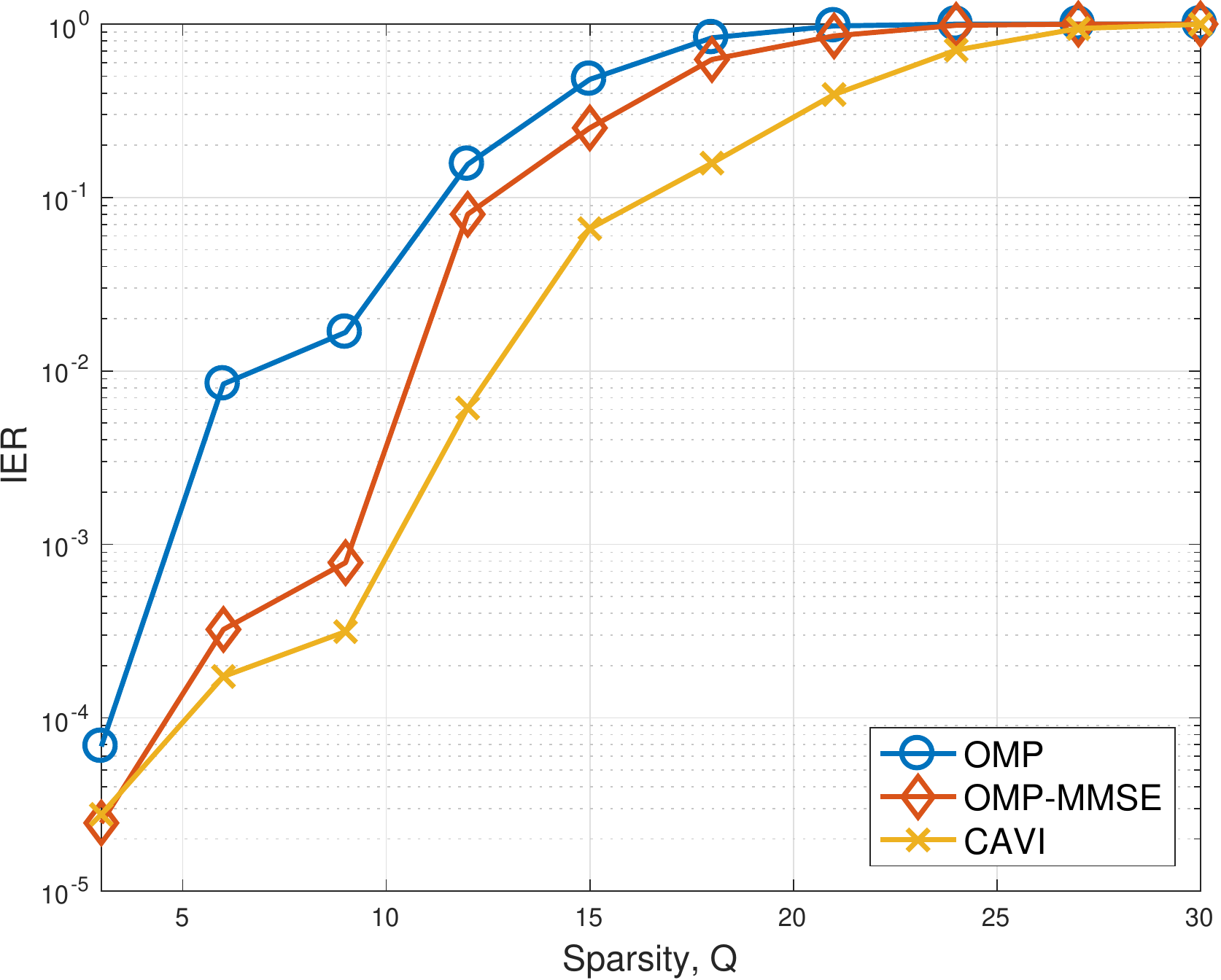} \\
(a) \\
\includegraphics[width=\figwidth]{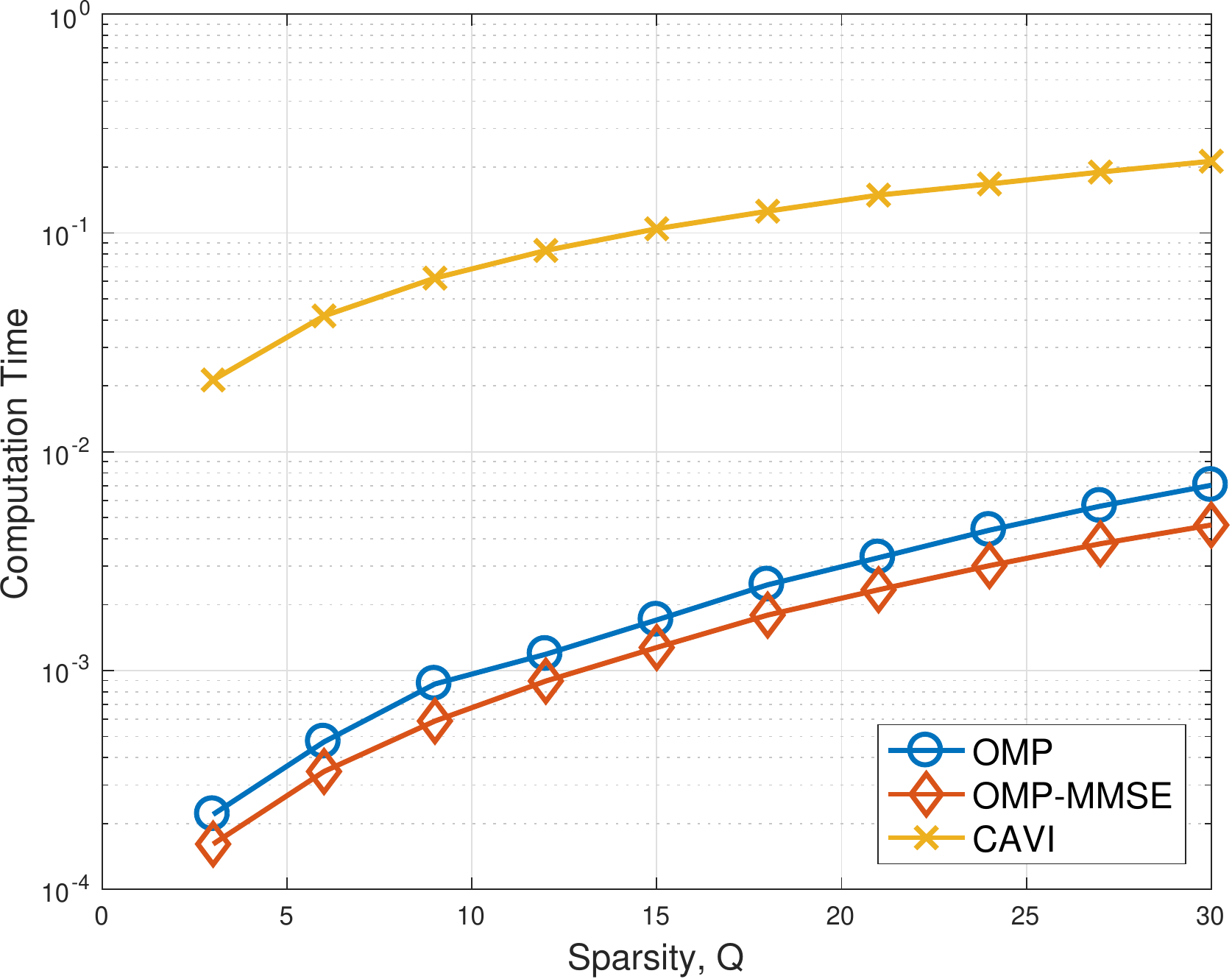} \\
(b) \\
\end{center}
\caption{IER and computation time 
of the 3 different detectors as functions of $Q$ (or $N$)
with fixed $D = 8$
when $L = 64$, $P = 4$, and SNR $= 16$dB: (a) IER; (b) computation time
(obtained by ``tic" and ``toc" commands of MATLAB).}
        \label{Fig:plt3}
\end{figure}

For a fixed $L$, SCIM with precoding can 
transmit more bits as $N$ increases. In particular, 
with a fixed $D$, $\tilde N_{\rm im}$ 
can grow linearly with 
$N$ or $Q$ as $\tilde N_{\rm im} = Q \log_2 D = \frac{N}{D} \log_2 D$.
In Fig.~\ref{Fig:plt3} (a),
we present the IER as a function of
sparsity $Q$ (or $N$) with fixed $D = 8$
when $L = 64$, $P = 4$, and SNR $= 16$dB, where 
a trade-off between the number of bits transmitted by IM (i.e.,
$\tilde N_{\rm im}$) and the performance of IER is shown.
That is, the increase of $\tilde N_{\rm im}$
results in a poor IER performance.
We note that the performance of the OMP-MMSE detector
is close to that of the CAVI detector when $Q$ is small.
However, as $Q$ increases, the performance of the OMP-MMSE detector
approaches that of the OMP detector.
Thus, when a sufficiently low IER is 
desirable with a small $Q$, the OMP-MMSE detector
(instead of  the CAVI detector)
can be used as it can provide a good performance with low-complexity.

In Fig.~\ref{Fig:plt3} (b), we show the computation times of
the 3 different detectors with the same parameter set as those used in
Fig.~\ref{Fig:plt3} (a). It is shown that
the computation time increases with $Q$ or $N$ and
the computation time of the CAVI detector 
is much higher than those of the OMP and OMP-MMSE detectors
as expected. In particular,
as mentioned earlier, 
the complexity order
of the CAVI algorithm is $L N_{\rm run}$ times higher than 
that of the OMP algorithm, so 
the CAVI detector requires a higher computation time 
when comparing the OMP detector by $L N_{\rm times} = 256$ times
or a factor of $10^2$, which is clearly shown in Fig.~\ref{Fig:plt3} (b).

Note that in  Fig.~\ref{Fig:plt3} (b),
the computation time of the OMP detector 
is higher than that of the OMP-MMSE detector.
This is due to the different measurement matrices
in \eqref{EQ:CSF} and \eqref{EQ:CSF_v}. 
In \eqref{EQ:CSF_v}, $\bPsi$ with FTN signaling has a number of zeros,
which allows efficient pseudo-inverse operations in the OMP 
algorithm and results in a low computation time.
Thus, the OMP-MMSE detector is better than
the OMP detector in terms of performance as well as computational complexity.

\begin{figure}[thb]
\begin{center}
\includegraphics[width=\figwidth]{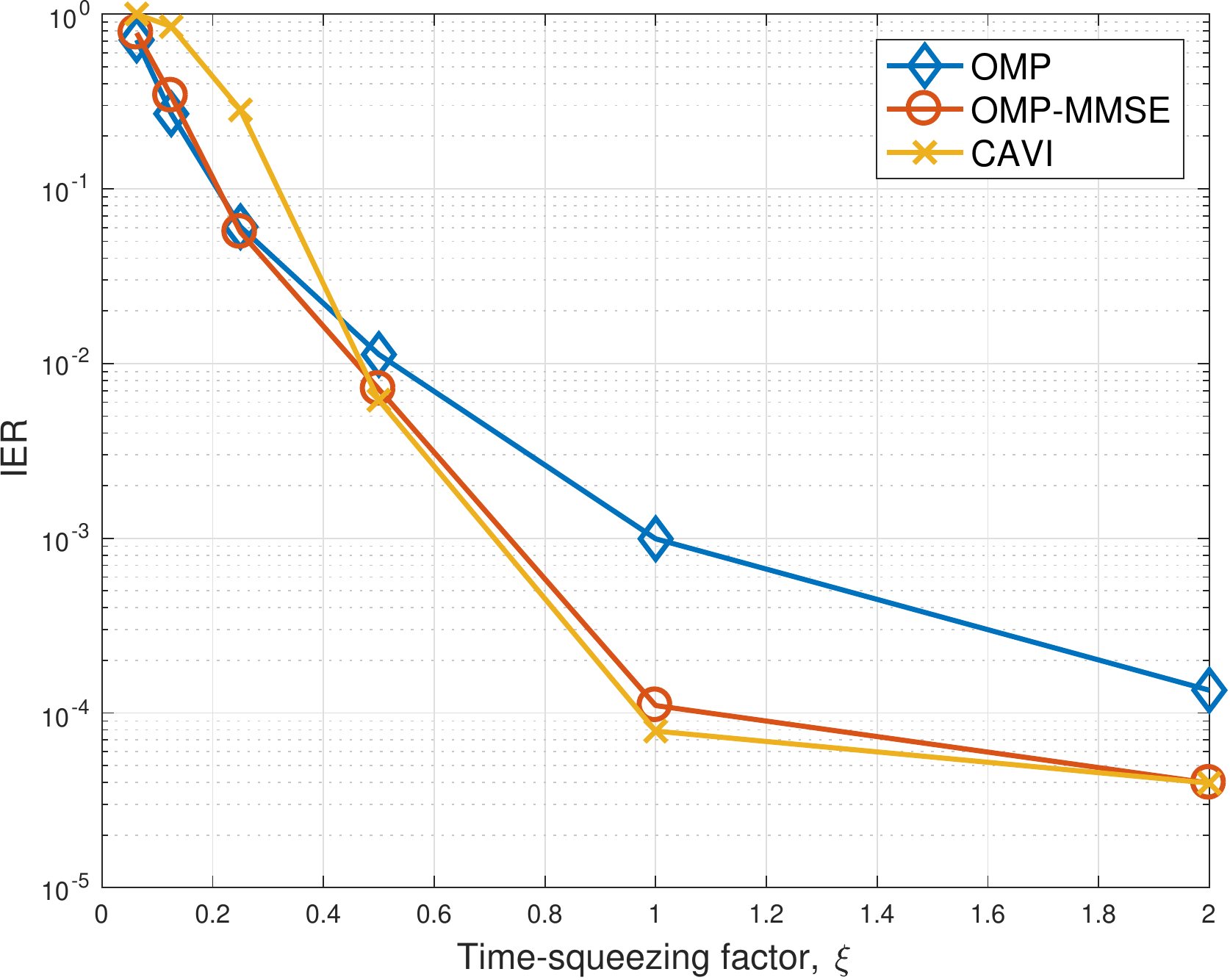} \\
\end{center}
\caption{IER as a function of the 
time-squeezing factor $\xi$ 
with $Q = 4$, $L = 64$, $P = 4$, and SNR $= 16$dB.}
        \label{Fig:plt6}
\end{figure}

With FTN precoding, the time-squeezing factor, $\xi$, 
affects performance.
To see this, we show the IER as a function of $\xi$
in Fig.~\ref{Fig:plt6} with $Q = 4$, $L = 64$, $P = 4$, and SNR $= 16$dB. 
Note that since
$Q$ is fixed, as $\xi$ increases, $N$ or $D$ also increases,
which also results in the increase of $\tilde N_{\rm im}$.
It is shown that the time-squeezing factor cannot be too small
to provide a reasonable performance.
For example, with $\xi = 0.5$, an IER of $10^{-2}$ can be achieved.
However, if $\xi$ becomes smaller than $0.5$,
the IER becomes too high.

\begin{figure}[thb]
\begin{center}
\includegraphics[width=\figwidth]{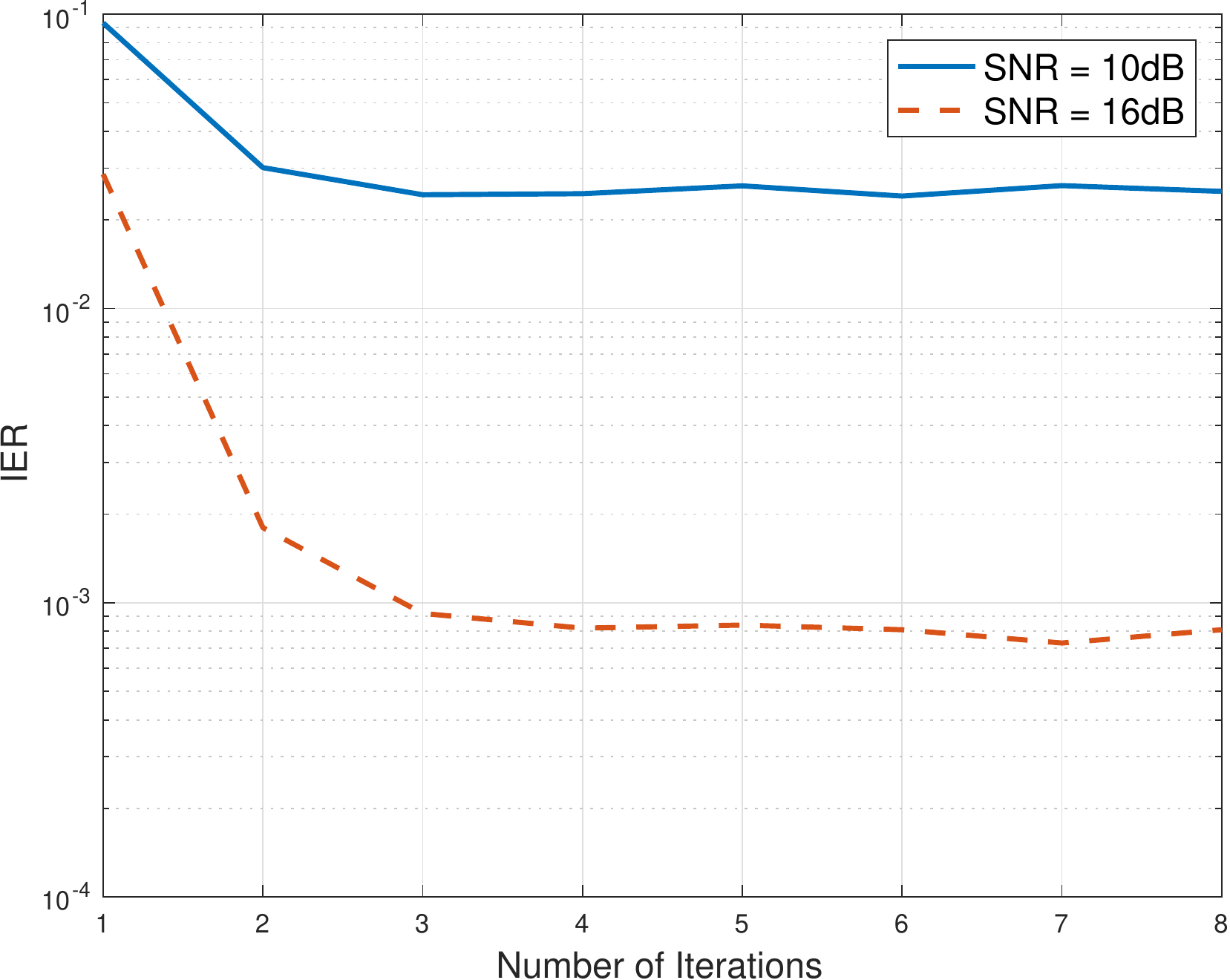} 
\end{center}
\caption{IER of CAVI as a function of the number of iterations,
$N_{\rm run}$ when $L = 64$, $P = 4$, $K = 2$, $(Q,D) = (5, 8)$,
and SNR $= 16$dB.}
        \label{Fig:Aplt0}
\end{figure}

We now consider the use of SCIM 
with FTN precoding to support multiple low-rate devices
in the same radio resource block.
As mentioned earlier, in this case, the OMP-MMSE detector
cannot be used.  
Thus, for a good performance with a reasonable computational complexity, 
we may need to use the CAVI detector.
Since the CAVI detector is based on an iterative algorithm,
its performance depends on the number of iterations, $N_{\rm run}$.
In Fig.~\ref{Fig:Aplt0}, we show
the IER of the CAVI detector as a function of the number of iterations,
$N_{\rm run}$, when $L = 64$, $P = 4$, $K = 2$, $(Q,D) = (5, 8)$,
and SNR $= 16$dB.
We can see that 3 or 4 iterations are sufficient for
the convergence at a medium or high SNR. 
Thus, as before, $N_{\rm run}$ is set to 4
in the rest of simulations.

\begin{figure}[thb]
\begin{center}
\includegraphics[width=\figwidth]{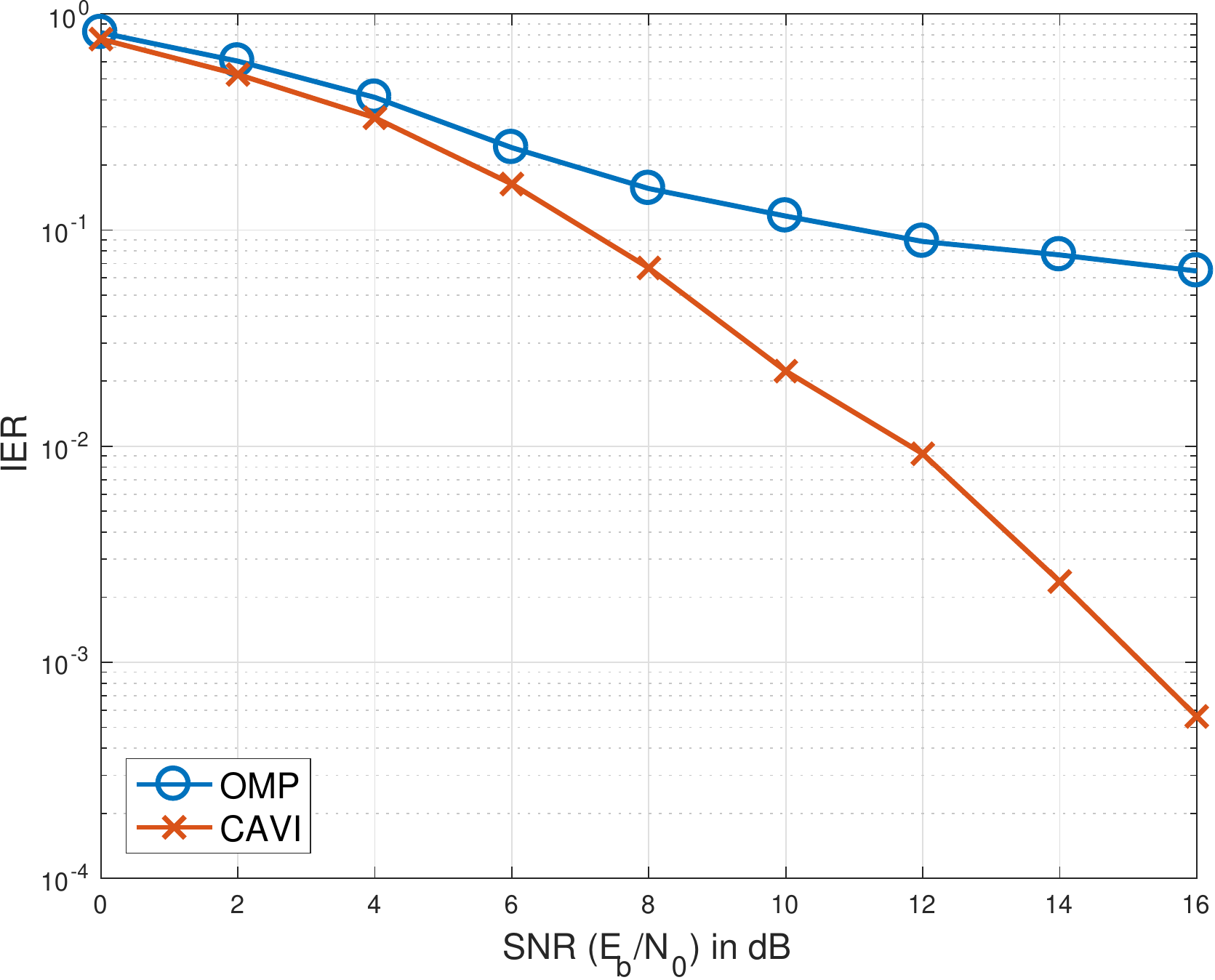} \\
\end{center}
\caption{IER of SCIM with multiple devices as a function of SNR 
when $L = 64$, $P = 4$, $Q = 5$, $D = 8$, and $K = 2$.}
        \label{Fig:Aplt1}
\end{figure}

In Fig.~\ref{Fig:Aplt1},
we show
the IER of SCIM with multiple devices as a function of SNR 
when $L = 64$, $P = 4$, $Q = 5$, $D = 8$, and $K = 2$.
In this case, the transmission rate
at each device is lower than the Nyquist rate at the receiver by
a factor of $40/64 = 0.625$, and each device has $\tilde N_{\rm im}
= 5\log_2 8 = 15$ bits to be transmitted by IM.
In addition, since 4-QAM is used for active symbols,
the total number of bits per block per device 
is $15 + 5 \log_2 4 = 25$ bits.
At an SNR of 16dB, the CAVI detector can provide an IER lower than 
$10^{-3}$ and the OMP detector can achieve an IER slightly lower
than $10^{-1}$. Clearly, it demonstrates that 
the simple OMP detector cannot be used,
but a more complicated detector, e.g., the CAVI detector,
is required to achieve a good performance.

\begin{figure}[thb]
\begin{center}
\includegraphics[width=\figwidth]{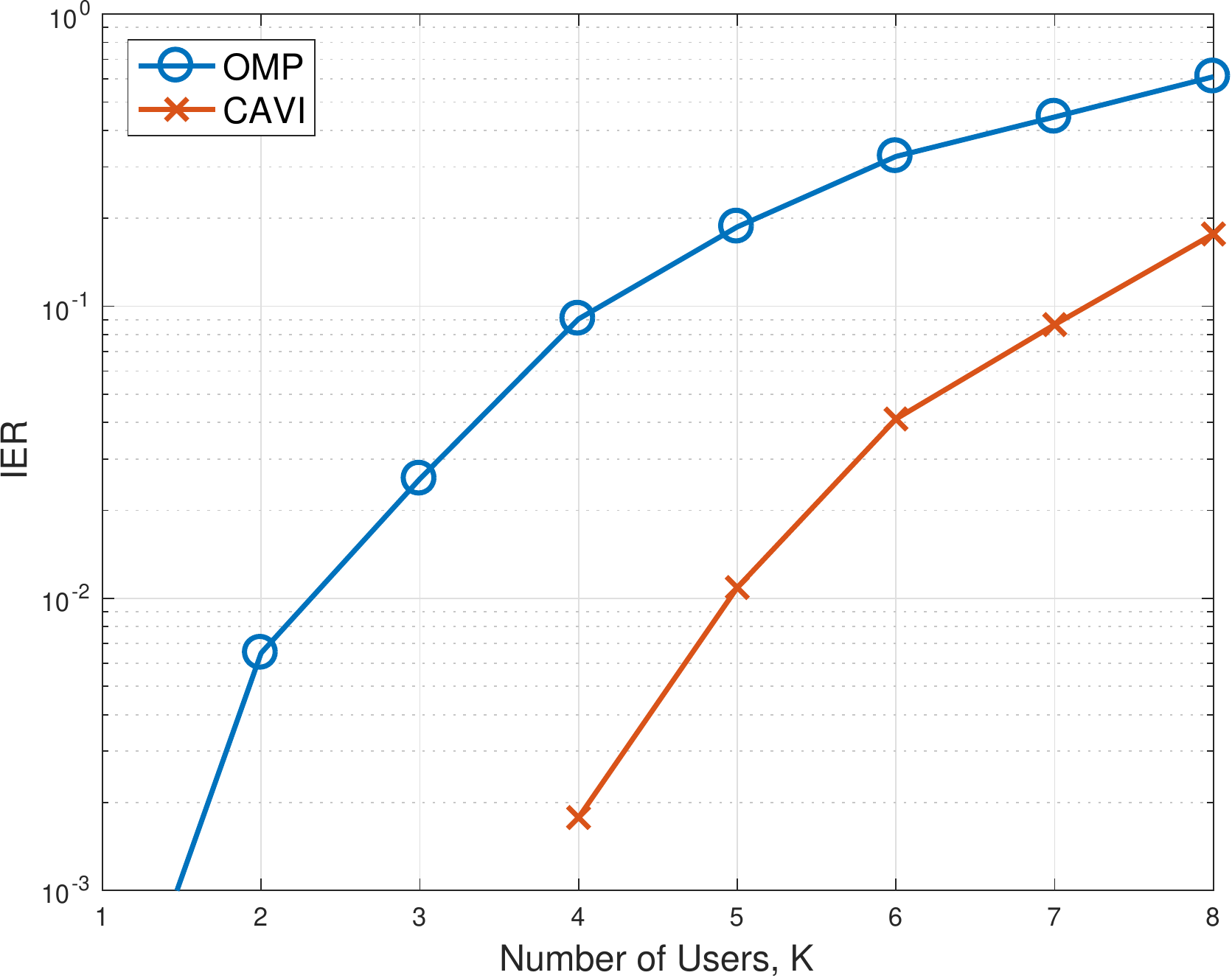} \\
\end{center}
\caption{IER of SCIM with multiple devices as a function of $K$
when $L = 64$, $P = 4$, $Q = 3$, $D = 8$, and SNR $= 16$ dB.}
        \label{Fig:Aplt2}
\end{figure}

Fig.~\ref{Fig:Aplt2}
shows 
the IER of SCIM with multiple devices as a function of $K$
when $L = 64$, $P = 4$, $Q = 3$, $D = 8$, and SNR $= 16$ dB.
Since $D$ and $Q$ are fixed, each device 
can transmit $\tilde N_{\rm im} = 3 \log_2 8 = 9$ bits by IM
regardless of $K$,
and the transmission rate
at each device is lower than the Nyquist rate at the receiver by
a factor of $24/64 = 0.375$ (since $N = QD = 24$).
It is shown that
as $K$ increases, 
the IER increases, which demonstrates a trade-off 
between the performance and 
the number of devices to be supported.
Clearly, 
the number of devices, $K$,
is to be limited for a reasonably good performance in terms of IER.
For example, at a target IER of $10^{-2}$,
$K$ can be up to 5 if the CAVI detector is used.

\begin{figure}[thb]
\begin{center}
\includegraphics[width=\figwidth]{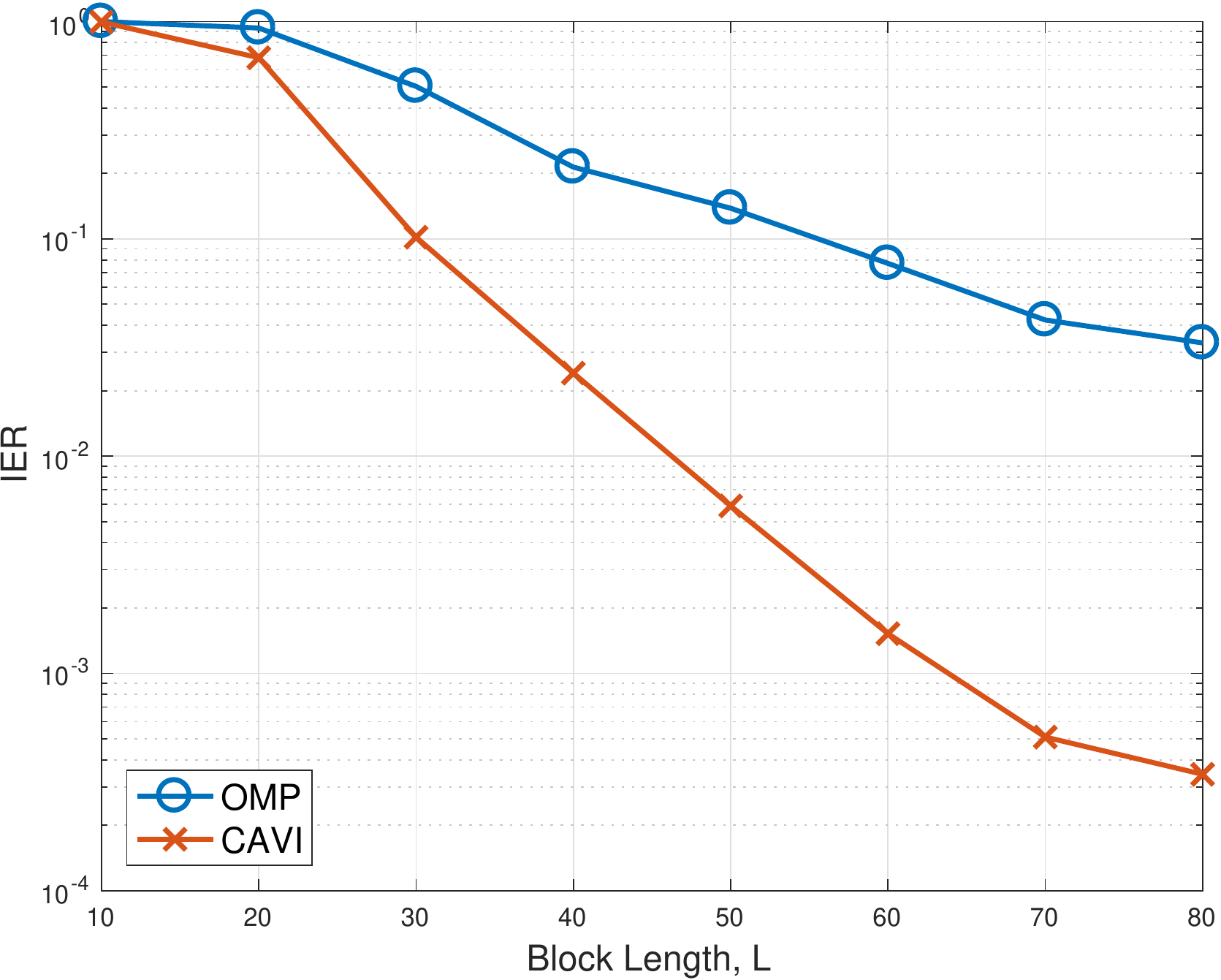} \\
\end{center}
\caption{IER of SCIM with multiple devices as a function of $L$
when $K = 2$, $P = 4$, $Q = 5$, $D = 8$, and SNR $= 16$ dB.}
        \label{Fig:Aplt4}
\end{figure}

The impact of $L$ on the performance
of SCIM with multiple devices in terms of IER 
is shown in Fig.~\ref{Fig:Aplt4}
when $K = 2$, $P = 4$, $Q = 5$, $D = 8$, and SNR $= 16$ dB.
Since $Q$ and $D$ are fixed, $N = 40$ is also fixed at each device.
Thus, as $L$ increases, we can assume that the system bandwidth
increases and the spectral efficiency decreases.
Thus, in Fig.~\ref{Fig:Aplt4}, we can see that
the IER decreases at the cost of
spectral efficiency
(i.e., lowering the spectral efficiency results in 
a lower IER).

\begin{figure}[thb]
\begin{center}
\includegraphics[width=\figwidth]{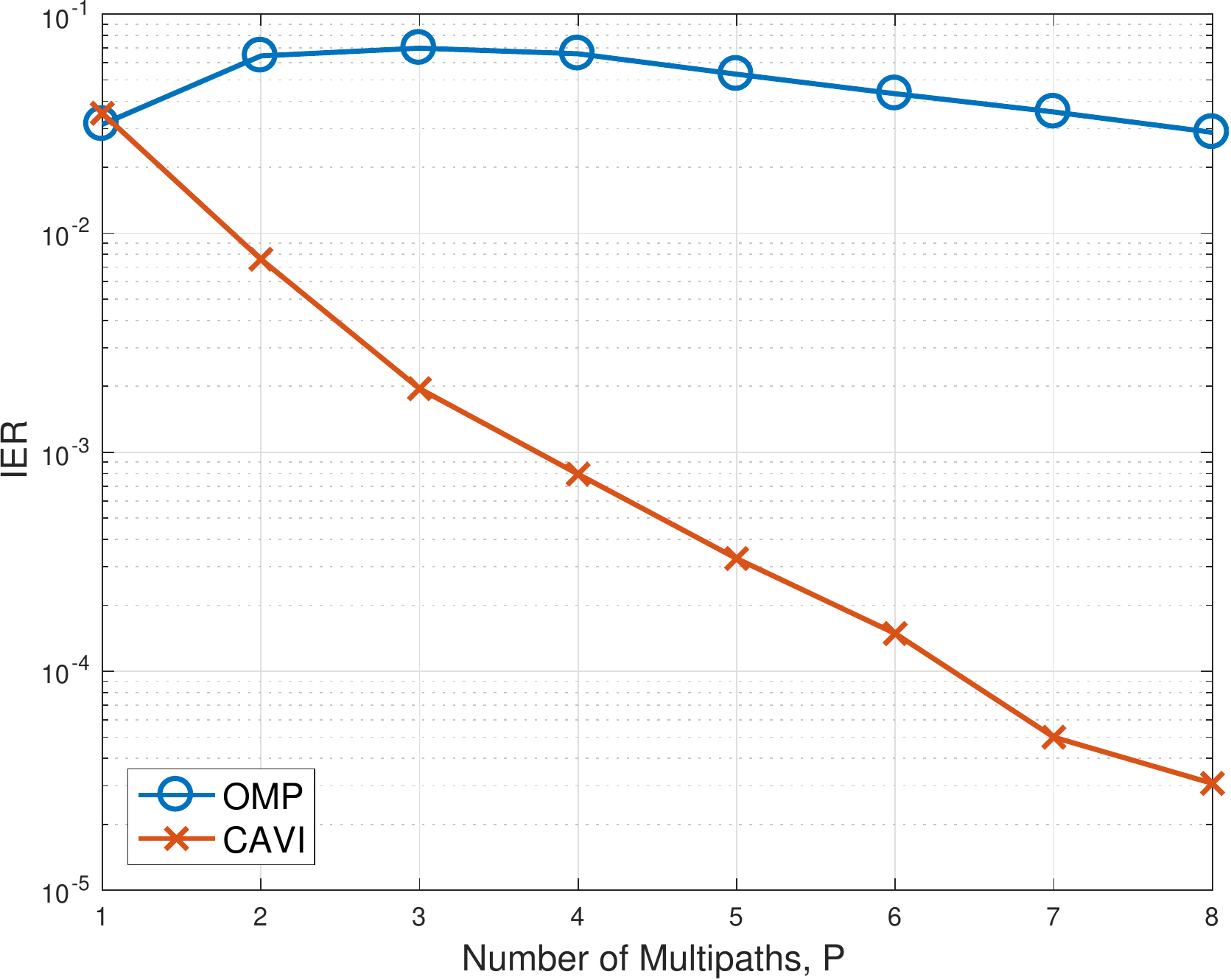} \\
\end{center}
\caption{IER of SCIM with multiple devices as a function of $P$
when $L = 64$, $K = 2$, $Q = 5$, $D = 8$, and SNR $= 16$ dB.}
        \label{Fig:Aplt3}
\end{figure}

As mentioned earlier,
an advantage of SCIM over OFDM-IM is the path diversity gain.
Thus, it is expected to see a better performance as $P$ increases.
In Fig.~\ref{Fig:Aplt3},
the IER is shown as a function of the number of multipaths, $P$,
with $L = 64$, $K = 2$, $Q = 5$, $D = 8$, and SNR $= 16$ dB.
As expected, due to a higher path diversity gain,
a lower IER can be achieved with increasing $P$ when the CAVI detector 
is used.
On the other hand, the OMP detector does not provide
an improved performance as $P$ increases.
This demonstrates that in order to fully exploit the path
diversity gain, an optimal detector or an approximate
optimal detector has to be used.

\section{Concluding Remarks}	\label{S:Conc}

In this paper, we studied SCIM
that is an application of IM to SC systems for
IoT uplink as it has several advantages, e.g., low PAPR,
path diversity gain, no inverse FFT operation
required, compared to OFDM-IM.
To increase the number of information bits transmitted by IM,
precoding was applied to SCIM.
In particular, FTN precoding has been considered.
We also generalized SCIM with precoding to multiple access channel
so that multiple devices can share the same radio
resource block for uplink transmissions.
With FTN precoding, we showed that devices can have
lower transmission rates than the receiver's sampling rate
(Nyquist rate), which might be 
useful when a device needs to lower its clock frequency 
for energy saving.
We also derived different detectors for sparse signal detection
including the CAVI detector that can provide an approximate
solution to the (optimal) MAP detection.

There are a number of further research works for SCIM with precoding.
For example, optimal precoding matrices can be 
designed under various constraints.
In particular, when devices have different transmission rates,
the design of optimal precoding matrices might be an interesting
and important topic. Channel coding can also be considered. 
Since the CAVI detector can provide soft-decisions, 
channel decoding can be performed with soft-decisions together
with the CAVI detector. Furthermore, in each iteration of the CAVI
algorithm, channel decoding can be performed, which can
result in a fast convergence rate as well as an improved performance.

\bibliographystyle{ieeetr}
\bibliography{cs}

\begin{thebibliography}{10}

\bibitem{Atzori10}
L.~Atzori, A.~Iera, and G.~Morabito, ``The {I}nternet of {T}hings: A survey,''
  {\em Comput. Netw.}, vol.~54, pp.~2787--2805, Oct. 2010.

\bibitem{ITU_IoT}
ITU-T, {\em Y.2060: Overview of the Internet of things}, June 2012.

\bibitem{Fuqaha15}
A.~Al-Fuqaha, M.~Guizani, M.~Mohammadi, M.~Aledhari, and M.~Ayyash, ``Internet
  of {T}hings: A survey on enabling technologies, protocols, and
  applications,'' {\em IEEE Communications Surveys Tutorials}, vol.~17,
  pp.~2347--2376, Fourthquarter 2015.

\bibitem{Sharma16}
S.~K. Sharma, T.~E. Bogale, S.~Chatzinotas, X.~Wang, and L.~B. Le, ``Physical
  layer aspects of wireless {IoT},'' in {\em 2016 International Symposium on
  Wireless Communication Systems (ISWCS)}, pp.~304--308, Sept 2016.

\bibitem{Gutierrez03}
J.~A. Gutierrez, E.~H. Callaway, and R.~Barrett, {\em IEEE 802.15.4 Low-Rate
  Wireless Personal Area Networks: Enabling Wireless Sensor Networks}.
\newblock New York, NY, USA: IEEE Standards Office, 2003.

\bibitem{3GPP_MTC}
3GPP TR 37.868 V11.0, {\em Study on {RAN} improvments for machine-type
  communications}, October 2011.

\bibitem{3GPP_NBIoT}
3GPP TS 36.321 V13.2.0, {\em Evolved Universal Terrestrial Radio Access
  ({E-UTRA}); Medium Access Control ({MAC}) protocol specification}, June 2016.

\bibitem{Wang17}
Y.~.~E. Wang, X.~Lin, A.~Adhikary, A.~Grovlen, Y.~Sui, Y.~Blankenship,
  J.~Bergman, and H.~S. Razaghi, ``A primer on {3GPP} narrowband {I}nternet of
  {T}hings,'' {\em IEEE Communications Magazine}, vol.~55, pp.~117--123, March
  2017.

\bibitem{Dahlman13}
E.~Dahlman, S.~Parkvall, and J.~Skold, {\em 4G: LTE/LTE-Advanced for Mobile
  Broadband, 2nd Edition}.
\newblock Academic Press, 2013.

\bibitem{Basar13}
E.~Basar, U.~Aygolu, E.~Panayirci, and H.~Poor, ``Orthogonal frequency division
  multiplexing with index modulation,'' {\em IEEE Trans. Signal Processing},
  vol.~61, pp.~5536--5549, Nov 2013.

\bibitem{Abu09}
R.~Abu-alhiga and H.~Haas, ``Subcarrier-index modulation {OFDM},'' in {\em
  Personal, Indoor and Mobile Radio Communications, 2009 IEEE 20th
  International Symposium on}, pp.~177--181, Sept 2009.

\bibitem{Xiao14b}
L.~Xiao, B.~Xu, H.~Bai, Y.~Xiao, X.~Lei, and S.~Li, ``Performance evaluation in
  {PAPR} and {ICI} for {ISIM-OFDM} systems,'' in {\em 2014 International
  Workshop on High Mobility Wireless Communications}, pp.~84--88, Nov 2014.

\bibitem{Fan15}
R.~Fan, Y.~J. Yu, and Y.~L. Guan, ``Generalization of orthogonal frequency
  division multiplexing with index modulation,'' {\em IEEE Trans. Wireless
  Communications}, vol.~14, pp.~5350--5359, Oct 2015.

\bibitem{Ko14}
Y.~Ko, ``A tight upper bound on bit error rate of joint {OFDM} and
  multi-carrier index keying,'' {\em IEEE Communications Letters}, vol.~18,
  pp.~1763--1766, Oct 2014.

\bibitem{Mesleh08}
R.~Y. Mesleh, H.~Hass, S.~Sinanovic, C.~W. Ahn, and S.~Yun, ``Spatial
  modulation,'' {\em IEEE Trans. Veh. Technol.}, vol.~57, pp.~2228--2241, July
  2008.

\bibitem{Jeganathan08}
J.~Jeganathan, A.~Ghrayeb, and L.~Szczecinski, ``Spatial modulation: optimal
  detection and performance analysis,'' {\em IEEE Commun. Letters}, vol.~12,
  pp.~545--547, Aug. 2008.

\bibitem{Renzo11}
M.~D. Renzo, H.~Haas, and P.~M. Grant, ``Spatial modulation for
  multiple-antenna wireless systems: a survey,'' {\em IEEE Communications
  Magazine}, vol.~49, pp.~182--191, December 2011.

\bibitem{Basar16}
E.~Basar, ``Index modulation techniques for {5G} wireless networks,'' {\em IEEE
  Communications Magazine}, vol.~54, pp.~168--175, July 2016.

\bibitem{Choi_Ko15}
J.~Choi and Y.~Ko, ``Compressive sensing based detector for sparse signal
  modulation in precoded {OFDM},'' in {\em Proc. IEEE ICC}, pp.~4536--4540,
  June 2015.

\bibitem{JChoi15}
J.~Choi, ``Sparse index multiple access,'' in {\em 2015 IEEE Global Conference
  on Signal and Information Processing (GlobalSIP)}, pp.~324--327, Dec 2015.

\bibitem{Choi16}
J.~Choi, ``Sparse index multiple access for multi-carrier systems with
  precoding,'' {\em J. Communications and Neworks}, vol.~18, pp.~3226--3237,
  June 2016.

\bibitem{Donoho06}
D.~Donoho, ``Compressed sensing,'' {\em IEEE Trans. Information Theory},
  vol.~52, pp.~1289--1306, April 2006.

\bibitem{Candes06}
E.~Candes, J.~Romberg, and T.~Tao, ``Robust uncertainty principles: exact
  signal reconstruction from highly incomplete frequency information,'' {\em
  IEEE Trans. Information Theory}, vol.~52, pp.~489--509, Feb 2006.

\bibitem{Basar15}
E.~{Basar}, ``Multiple-input multiple-output {OFDM} with index modulation,''
  {\em IEEE Signal Processing Letters}, vol.~22, pp.~2259--2263, Dec 2015.

\bibitem{Zheng17}
B.~{Zheng}, M.~{Wen}, E.~{Basar}, and F.~{Chen}, ``Multiple-input
  multiple-output ofdm with index modulation: Low-complexity detector design,''
  {\em IEEE Trans. Signal Processing}, vol.~65, pp.~2758--2772, June 2017.

\bibitem{Gao18}
S.~{Gao}, M.~{Zhang}, and X.~{Cheng}, ``Precoded index modulation for
  multi-input multi-output ofdm,'' {\em IEEE Trans. Wireless Communications},
  vol.~17, pp.~17--28, Jan 2018.

\bibitem{Basar15_CI}
E.~{Basar}, ``{OFDM} with index modulation using coordinate interleaving,''
  {\em IEEE Wireless Communications Letters}, vol.~4, pp.~381--384, Aug 2015.

\bibitem{Liu17}
Y.~{Liu}, F.~{Ji}, H.~{Yu}, F.~{Chen}, D.~{Wan}, and B.~{Zheng}, ``Enhanced
  coordinate interleaved {OFDM} with index modulation,'' {\em IEEE Access},
  vol.~5, pp.~27504--27513, 2017.

\bibitem{Li18}
Q.~{Li}, M.~{Wen}, E.~{Basar}, H.~V. {Poor}, B.~{Zheng}, and F.~{Chen},
  ``Diversity enhancing multiple-mode {OFDM} with index modulation,'' {\em IEEE
  Trans. Communications}, vol.~66, pp.~3653--3666, Aug 2018.

\bibitem{Choi17_CIM}
J.~{Choi}, ``Coded {OFDM-IM} with transmit diversity,'' {\em IEEE Trans.
  Communications}, vol.~65, pp.~3164--3171, July 2017.

\bibitem{Choi18_TCM}
J.~{Choi} and Y.~{Ko}, ``{TCM} for {OFDM-IM},'' {\em IEEE Wireless
  Communications Letters}, vol.~7, pp.~50--53, Feb 2018.

\bibitem{Sugiura17}
S.~Sugiura, T.~Ishihara, and M.~Nakao, ``State-of-the-art design of index
  modulation in the space, time, and frequency domains: Benefits and
  fundamental limitations,'' {\em IEEE Access}, vol.~5, pp.~21774--21790, 2017.

\bibitem{JC_ICC17}
J.~Choi, ``Single-carrier index modulation and {CS} detection,'' in {\em 2017
  IEEE International Conference on Communications (ICC)}, pp.~1--6, May 2017.

\bibitem{Nakao17}
M.~Nakao, T.~Ishihara, and S.~Sugiura, ``Single-carrier frequency-domain
  equalization with index modulation,'' {\em IEEE Communications Letters},
  vol.~21, pp.~298--301, Feb 2017.

\bibitem{Falconer02}
D.~Falconer, S.~L. Ariyavisitakul, A.~Benyamin-Seeyar, and B.~Eidson,
  ``Frequency domain equalization for single-carrier broadband wireless
  systems,'' {\em IEEE Communications Magazine}, vol.~40, pp.~58--66, Apr 2002.

\bibitem{Mazo75}
J.~E. Mazo, ``Faster-than-{N}yquist signaling,'' {\em The Bell System Technical
  Journal}, vol.~54, pp.~1451--1462, Oct 1975.

\bibitem{Ishihara17}
T.~Ishihara and S.~Sugiura, ``{F}aster-than-{N}yquist signaling with index
  modulation,'' {\em IEEE Wireless Communications Letters}, vol.~6,
  pp.~630--633, Oct 2017.

\bibitem{Jordan99}
M.~I. Jordan, Z.~Ghahramani, T.~S. Jaakkola, and L.~K. Saul, ``An introduction
  to variational methods for graphical models,'' {\em Machine Learning},
  vol.~37, pp.~183--233, Nov 1999.

\bibitem{Spencer14}
J.~Spencer, {\em Asymptopia}.
\newblock American Mathematical Society, 2014.

\bibitem{ChoiJBook}
J.~Choi, {\em Adaptive and Iterative Signal Processing in Communications}.
\newblock Cambridge University Press, 2006.

\bibitem{Fan17}
J.~Fan, S.~Guo, X.~Zhou, Y.~Ren, G.~Y. Li, and X.~Chen, ``Faster-than-nyquist
  signaling: An overview,'' {\em IEEE Access}, vol.~5, pp.~1925--1940, 2017.

\bibitem{Anderson13}
J.~B. Anderson, F.~Rusek, and V.~Öwall, ``Faster-than-{N}yquist signaling,''
  {\em Proceedings of the IEEE}, vol.~101, pp.~1817--1830, Aug 2013.

\bibitem{Pancaldi08}
F.~Pancaldi, G.~M. Vitetta, R.~Kalbasi, N.~Al-Dhahir, M.~Uysal, and H.~Mheidat,
  ``Single-carrier frequency domain equalization,'' {\em IEEE Signal Processing
  Magazine}, vol.~25, pp.~37--56, September 2008.

\bibitem{Candes08}
E.~Candes and M.~B. Wakin, ``An introduction to compressive sampling,'' {\em
  IEEE Signal Processing Magazine}, vol.~25, no.~2, pp.~21--30, 2008.

\bibitem{Eldar12}
Y.~C. Eldar and G.~Kutyniok, {\em Compressed Sensing: Theory and Applications}.
\newblock Cambridge University Press, 2012.

\bibitem{Candes08b}
E.~J. Candes, ``The restricted isometry property and its implications for
  compressed sensing,'' {\em Comptes Rendus Mathematique}, vol.~346,
  no.~9–10, pp.~589--592, 2008.

\bibitem{Baraniuk08}
R.~Baraniuk, M.~Davenport, R.~DeVore, and M.~Wakin, ``A simple proof of the
  restricted isometry property for random matrices,'' {\em Constructive
  Approximation}, vol.~28, no.~3, pp.~253--263, 2008.

\bibitem{Pati93}
Y.~Pati, R.~Rezaiifar, and P.~Krishnaprasad, ``Orthogonal matching pursuit:
  recursive function approximation with applications to wavelet
  decomposition,'' in {\em Signals, Systems and Computers, 1993. 1993
  Conference Record of The Twenty-Seventh Asilomar Conference on}, pp.~40--44
  vol.1, Nov 1993.

\bibitem{Davis94}
G.~M. Davis, S.~G. Mallat, and Z.~Zhang, ``Adaptive time-frequency
  decompositions,'' {\em Optical Engineering}, vol.~33, no.~7, pp.~2183--2191,
  1994.

\bibitem{Bishop06}
C.~M. Bishop, {\em Pattern Recognition and Machine Learning (Information
  Science and Statistics)}.
\newblock Berlin, Heidelberg: Springer-Verlag, 2006.

\bibitem{ChoiJBook2}
J.~Choi, {\em Optimal Combining and Detection}.
\newblock Cambridge University Press, 2010.

\bibitem{Blei17}
D.~M. Blei, A.~Kucukelbir, and J.~D. McAuliffe, ``Variational inference: A
  review for statisticians,'' {\em Journal of the American Statistical
  Association}, vol.~112, no.~518, pp.~859--877, 2017.

\bibitem{CoverBook}
T.~M. Cover and J.~A. Thomas, {\em Elements of Inform. Theory}.
\newblock NJ: John Wiley, second~ed., 2006.

\bibitem{Choi_VI18}
J.~{Choi}, ``A variational inference based detection method for repetition
  coded generalized spatial modulation,'' {\em IEEE Trans. Communications},
  pp.~1--1, 2018 (to be published).

\end{thebibliography}

\end{document}